\documentclass[%
  reprint,            % use "preprint" for single-column draft
  aps,                % American Physical Society
  prd,                % target journal: Physical Review D
  superscriptaddress, % common for multi-institution papers
  nofootinbib         % optional: move footnotes into text
]{revtex4-2}

\usepackage{graphicx}% Include figure files
\usepackage{dcolumn}% Align table columns on decimal point
\usepackage{bm}% bold math
\usepackage{lipsum}
\usepackage{amsmath}
\usepackage{amssymb}

\usepackage[dvipsnames]{xcolor} \usepackage[colorlinks=true]{hyperref}

\hypersetup{
    colorlinks=true,       
    citecolor=Periwinkle, % Color for \cite{} 
    linkcolor=Rhodamine,    % Color for \ref{} 
    urlcolor=Periwinkle  % Color for web URLs
}

\begin{document}

\preprint{APS/123-QED}

\title{Emulating Cosmic Structure Formation \\ with a Lagrangian Neural Cellular Automaton}

\author{Cooper Jacobus}
\email{jacobus@stanford.edu}
\affiliation{Department of Physics, Stanford University, Stanford, CA 94305, USA}
\affiliation{Kavli Institute for Particle Astrophysics and Cosmology, Stanford}

\author{Beatriz Tucci}
\affiliation{Kavli Institute for Particle Astrophysics and Cosmology, Stanford}
\affiliation{Leinweber Institute for Theoretical Physics, Stanford}

\author{Oliver H.\,E. Philcox}
\affiliation{Department of Physics, Stanford University, Stanford, CA 94305, USA}
\affiliation{Kavli Institute for Particle Astrophysics and Cosmology, Stanford}
\affiliation{Leinweber Institute for Theoretical Physics, Stanford}

\date{\today}

\begin{abstract}
\noindent
Field-level inference of cosmological initial conditions from galaxy surveys requires a forward model that is simultaneously accurate in the non-linear regime, computationally efficient, and fully differentiable. Traditional N-body simulations are accurate but computationally prohibitive for iterative inference, while approximate solvers like Lagrangian Perturbation Theory (LPT) fail to capture the knotty halo-forming dynamics of the cosmic web at late times. We introduce the \textit{Lagrangian Neural Cellular Automaton} (LNCA), a hybrid deep learning framework that can be applied to emulate structure formation as a local, iterative dynamical process on a comoving lattice. Unlike convolutional emulators which map fixed grids in a single pass, the LNCA operates in the Lagrangian frame, iteratively advecting the computational graph itself to follow the flow of mass. By training the network to learn only the \textit{residual} displacement corrections to the Zeldovich approximation, we achieve high-fidelity emulation of the non-linear physics while guaranteeing accuracy at large scales.
We further constrain our model to produce complete trajectories, not just final states, by adopting an equivariant cellular automaton architecture, which recurrently iterates on its internal states to yield a dynamic history.
The resulting update rule acts on a single fixed neighborhood, is translationally and rotationally equivariant, and is integrated over discrete substeps to yield trajectories, making it a differentiable forward model for reconstructing the initial conditions of the universe from lightcone data.
We train our model to emulate the real-space matter distributions of a diverse set of N-body simulations from the Quijote suite, for a range of redshifts and cosmologies, given only the initial conditions.
Our trained model supports percent-level precision in the power and cross spectra well into the non-linear regime ($k \lesssim 0.5 \, h \text{Mpc}^{-1}$), while requiring $\sim10^3$ times fewer learned parameters than comparable models and which take the form of an interpretable internal dynamic rule set.

\end{abstract}
%\oliver{I broadly like this abstract, but it's a bit vague on what we do, and a bit more of an intro than an abstract in some sense. We can be a bit more specific about what this paper actually does. e.g., we emulate real-space matter distribution in 3D, matching N-body sims. We get this \% accuracy at this scale. It's super fast. It's simple. It has much lower dimensionality than other methods. We use some cool techniques to train it. We could do other things in the future.}

\maketitle

\section{Introduction}
\label{sec:introduction}

\noindent The next generation of galaxy redshift surveys, including DESI \cite{DESI2016}, \textit{Euclid} \cite{Euclid2011}, \textit{SphereX} \cite{spherex}, \textit{Roman} \cite{roman, akeson2019}, and LSST \cite{LSST2009}, will map the distribution of matter in the Universe with unprecedented volume and precision. In this new data-rich era of cosmology, standard summary statistics such as the power spectrum or two-point correlation function are insufficient to extract the full physical information content of these maps. The non-linear and non-Gaussian nature of the universe's structure in its later epochs encourage a methodological transition toward \textit{field-level inference}, a paradigm where the non-linear fields and initial conditions are explicitly reconstructed such that they match the observed galaxy distribution \citep{Jasche2013, Wang2014, Seljak2017, McAlpine2025, Hahn2023, Nguyen2024}. Modeling the evolution of structure at the field-level itself will enable us to extract better constraints on cosmological physics by leveraging the enormous volumes of correlated information that standard summary statistic compressions necessarily throw out. However, this inverse problem requires a reliable forward model that connects linear (Gaussian) initial conditions of the early-universe to the observed late-time non-linear structures.

\newpage

Ideally, this forward model must satisfy three competing constraints: it must be physically accurate in the non-linear regime, it must be computationally cheap to allow for millions of sampling steps even when modeling incredibly large volumes, and it should be fully differentiable to enable gradient-based optimization \cite{Modi2021}.

Standard N-body simulations, while accurate and useful for inference problems on smaller volumes \citep{Hahn2023}, are too computationally expensive for the inner loop of a Bayesian inference pipeline of the requisite physical scale. Conversely, approximate solvers based on Lagrangian Perturbation Theory (LPT) \cite{Buchert1992, Bouchet1995} are highly efficient but break down in the non-linear regime as trajectories cross and structures virialize; nevertheless, LPT and Effective Field Theory (EFT) based forward models remain the backbone of many recent field-level inference pipelines, which either accept this regime of validity as sufficient for their target scales or supplement it with an emulated non-linear correction \citep{McAlpine2025, Doeser2024, Nusser2026, Nguyen2024, Babic2025}. A successful middle ground is the COmoving Lagrangian Acceleration (COLA) method \cite{Tassev2013}, which solves for the large-scale displacement analytically using LPT and integrates the residual non-linear forces numerically, and which has since been extended into several parallel, high-performance implementations \citep{Howlett2015, Feng2016}. However, standard COLA implementations rely on Particle-Mesh (PM) force calculations, which can be computationally demanding and difficult to differentiate efficiently at scale.

%\newpage

In parallel, deep learning has emerged as a particularly promising paradigm for cosmological emulation, driven by the ability of neural networks to approximate high-dimensional, non-linear mappings from large training 
sets of simulations \cite{Ntampaka2021}. 
Machine Learning methods are producing increasingly compelling surrogate models from power spectrum emulators to direct parameter inference \citep{Kwan2015, Modi2021}, and are being considered for implementation into modern survey analysis pipelines.
Many deterministic approaches apply convolutional networks to Lagrangian-space fields, where the displacement of a regular grid of fluid elements in treated as a three-dimensional image that standard U-Net and V-Net architectures process directly \cite{He2019, Jamieson2022, Kaushal2022}. Generative methods operating directly on Eulerian density fields can also be fruitful \cite{Ramanah2019, BergerStein2019}. While successful at texture synthesis, such CNNs typically lack Galilean invariance and struggle to model the advection of mass over large distances without prohibitively deep architectures, as the fixed, local receptive field of a convolutional kernel is poorly suited to representing bulk mass transport and breaks the discrete translation/permutation symmetries that make particle-based representations natural for this problem.

\begin{figure*}[t]
  \centering
  \includegraphics[width=\linewidth]{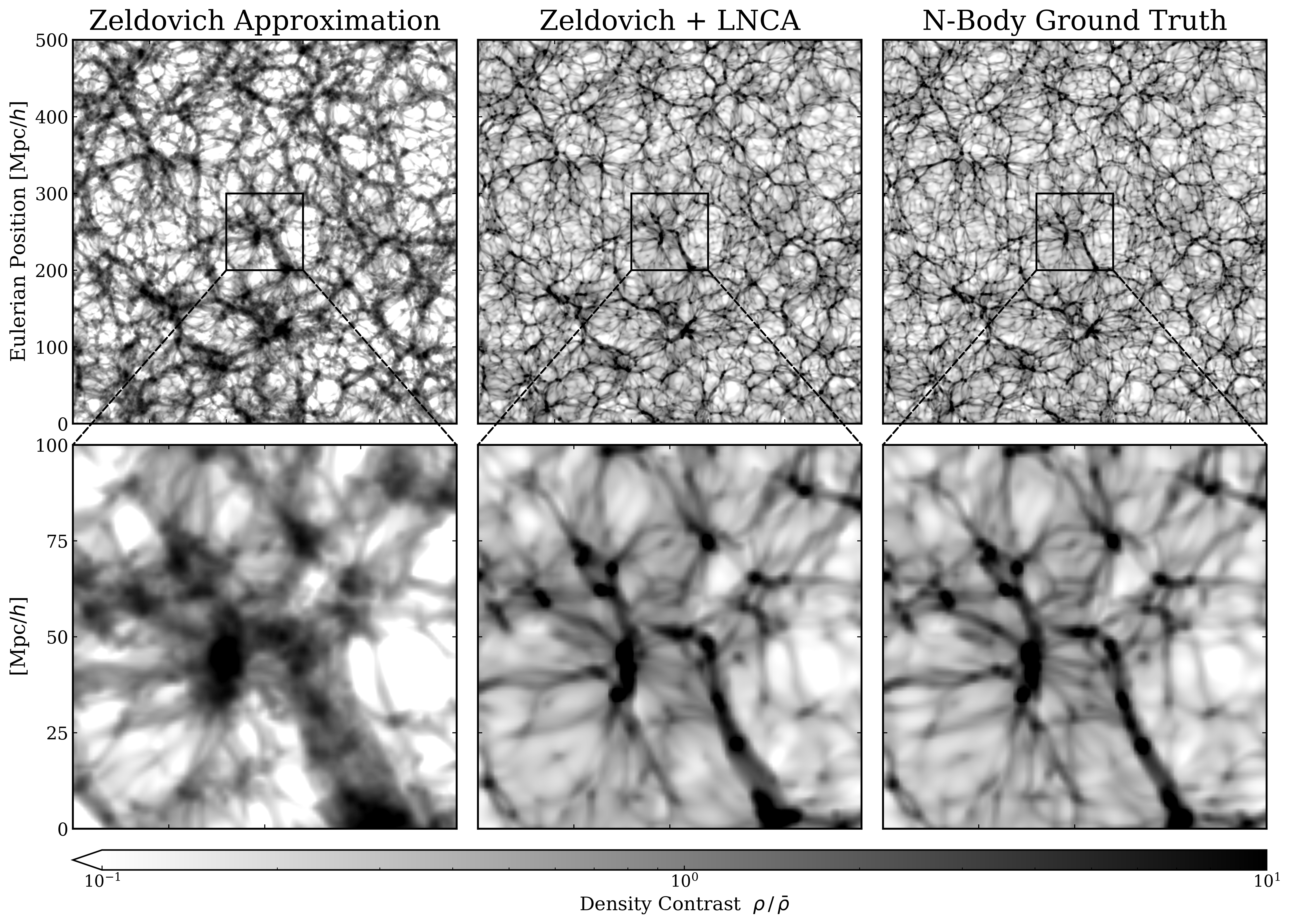}
  \vspace{-12pt}
  \caption{Comparison between 2D phase-sheet projections of the cosmic structure at redshift $z=0$ for: the Zeldovich Approximation alone (left), the Zeldovich Approximation corrected by our machine learning method (center), and a full N-body gravity simulation from the Quijote Suite (right) which was not seen by the model during training. 
  %Particles are colored according to the phase of their displacement direction to indicate motion and visually separate bulk flows.
  }
  \label{fig:boxes}
\end{figure*}

A broad class of approaches 
seeks to learn a direct mapping from linear initial conditions, which 
are efficiently tractable via linear perturbation theory, to the 
late-time non-linear density or displacement fields 
\citep[e.g.,][]{Jamieson2022, He2019, Ramanah2020, Dai2022, Zhang2024}. These models typically 
employ volumetric convolutional architectures like U-Nets, which efficiently encode multi-scale spatial correlations 
through hierarchical feature extraction and skip connections. By 
operating directly on three-dimensional fields rather than on 
particle catalogs, such architectures are well-suited to the 
statistical structure of the cosmic web, which exhibits coherent 
features spanning many orders of magnitude in spatial scale. 
Some newer models are trained to predict 
displacement fields rather than Eulerian density fields, offering the 
additional advantage of implicitly encoding mass conservation, 
mirroring the structure of Lagrangian perturbation theory 
\cite{Jamieson2025}. Our model belongs to this Lagrangian-displacement 
family, but departs from the single-pass U-Net mapping in a critical 
respect: rather than predicting the final displacement field in one 
forward evaluation, we recover it as the endpoint of an explicit, 
iterated integration over cosmic time, so that the same learned local 
rule is reused at every step of the flow.

A complementary class of approaches takes as its starting point a 
fast, approximate physical solver, such as the Zeldovich 
Approximation, \textsc{FastPM} \citep{Feng2016}, or \textsc{COLA} 
\citep{Tassev2013}, and uses machine 
learning to correct for the residual non-linearities that such 
solvers fail to capture \citep[e.g.,][]{Tosone2021, Chartier2021, 
Kaushal2022, Dai2024}. These backbones span a range of dynamical 
fidelity: the Zeldovich Approximation and \textsc{COLA} explicitly 
factor out low-order Lagrangian perturbation theory and leave the 
higher-order dynamics to be supplied elsewhere, whereas \textsc{FastPM} 
integrates the full particle-mesh forces and is approximate only in 
its coarse time-stepping and force resolution. This hybrid paradigm is motivated by the observation 
that much of the large-scale structure can be accounted for by a 
comparatively inexpensive dynamical core, and that the role of the neural 
network is therefore to act as a targeted \emph{subgrid correction} 
rather than a global function approximator. Such architectures tend 
to generalize more robustly across cosmological parameters, since the 
physics-informed backbone already encodes the dominant scaling 
behaviors, leaving the network to learn comparatively lower-amplitude 
residuals \cite{Chartier2021, Bartlett2025}. Our approach sits firmly within this 
tradition: we adopt the \textsc{COLA} decomposition, hard-coding the 
linear growth of the Zeldovich displacement and tasking the network 
only with the residual, non-linear correction. Where our model differs 
is that the correction is not applied as a one-shot post-processing 
step on a precomputed solver output, but is instead injected 
\emph{in the loop}, as a strictly local update evaluated at every 
substep of the integration. This lets the network sharpen collapsing 
features and correct the unphysical post-shell-crossing evolution of linear theory 
progressively, as structure forms, rather than repairing it after the 
fact.

Beyond direct field emulation, machine learning has also been applied 
to the inverse problem of reconstruction, inferring the linear 
initial conditions or cosmological parameters from observed 
large-scale structure \cite{Seljak2017, Modi2021, Doeser2024}. 
Field-level inference frameworks such as \textsc{borg} embed a differentiable forward model within a Bayesian hierarchical scheme and sample the posterior over initial conditions, and have recently been scaled to survey data to build detailed reconstructions of our local cosmic neighborhood \citep{Jasche2013, McAlpine2025, Porqueres2021, Nguyen2020}. Such methods 
place particularly stringent demands 
on the emulator: the model must be not only accurate but 
differentiable with respect to both the input fields and the 
cosmological parameters, so that adjoint gradients can be propagated 
back to the initial white-noise field. This 
requirement has spurred interest in architectures that couple neural 
networks with automatic differentiation frameworks such as 
\textsc{jax}, enabling end-to-end gradient 
propagation through the emulator for use in gradient-based samplers 
\citep{Doeser2024}. 

\vspace{12pt}

Despite these advances, several persistent challenges remain. 
Feed-forward architectures trained to map initial to final conditions 
in a single pass return one epoch per evaluation, so intermediate snapshots and lightcones require either a separate pass per epoch or interpolation between a few of them \cite{Jamieson2025}.  
Furthermore, purely data-driven 
models frequently struggle to extrapolate into regions of parameter 
space underrepresented in the training set, a limitation that 
physically motivated priors can help mitigate \cite{
Villaescusa-Navarro2021}. Finally, most architectures of all kinds above offer no way to inspect their inner representations or extract insight into the nature of their predictions.
The ideal emulator would be simultaneously 
accurate, fast, differentiable, temporally continuous, 
generalizable across cosmological parameters, and would yield to direct interpretation of its internal mechanics.

\newpage

In this work, we introduce a novel architecture designed specifically to address the above  
limitations, which we  call a \textit{Lagrangian Neural Cellular Automaton}. 
We demonstrate that the challenges of hybrid cosmological forward modeling can be simultaneously addressed by embedding an equivariant learned neural operator within a temporally continuous, Lagrangian physical integrator.
Rather than bypassing the physics entirely or correcting only the final output, 
our model applies iterative neural updates at each time step 
of a discretized Lagrangian flow, allowing the model to track 
particle displacement histories continuously from the initial conditions ($z=127$) to the final state ($z=0$).

%Traditional Cellular Automata (CAs) are discretized dynamical systems where global complexity emerges from simple, local update rules \cite{Mordvintsev2020} \oliver{older cites too? [or mention if this is a review]}. To apply this framework in our context, we interpret the cosmological density field as a regular Lagrangian lattice where each node represents a fluid element endowed with physical state variables (displacement, velocity) as well as learned internal properties. The ``update rule" is parameterized by a small, rotationally equivariant neural network that computes interactions within a strictly localized ``Moore" neighborhood. In contrast with traditional CAs, our method applies ``forces'' to the cells which are integrated to update their values, rather than updating them directly. Critically, we also allow the cells themselves to be displaced by these forces, affecting their relationships with (\textit{i.e.}\ distances to) their local neighbors. 

To allow our model to focus only on the difficult non-linear dynamics, our network does not learn the total gravitational force, but rather the \textit{residual} force required to correct the Zeldovich approximation. This enforces large-scale consistency while allowing the network to dedicate all of its expressive power to small-scale clustering, and enables the learned update rule to be \textit{strictly} localized post-LPT.

%\newpage

%To allow the model to accurately capture these non-linear dynamics, we construct it as a family of several interacting scalar and vector fields in Lagrangian space, where it learns the coupling behavior between these fields in an strictly equivariant basis. One of these vector fields is the Lagrangian displacement, which the model is trained to optimize relative to a suite of modern N-body simulations of varying cosmologies.

\newpage

The result is a recurrent, differentiable dynamical system that retains the parallelized computational efficiency of CNNs (via tensorized operations) while respecting the Lagrangian nature and symmetries of cosmic evolution. We demonstrate that our model can accurately recover the non-linear displacement field for a range of redshifts ($z \in [0,0.5,1,2,3]$), providing a fast, differentiable forward model for Bayesian reconstruction of the dark matter distribution and initial conditions given an observational catalog and a bias or halo model. %\oliver{good, but there are a few limitations: (1) it doesn't do baryons, (2) we'd need a galaxy-halo connection to actually model galaxies.}

\vspace{12pt}

The remainder of this paper is structured as follows: First we motivate and discuss the specific architecture of our model and outline the training strategy in section \S\ref{sec:model_design}. Then we present the results of our validation tests in \S\ref{sec:validation}. Finally, we discuss the implications of these results and the future utility of our model in section \S\ref{sec:conclusion}. Further training details and model constraints are discussed in the appendix.

\newpage

\section{Model Design}
\label{sec:model_design}

\noindent In the following section, we discuss the architecture and physical motivation of our proposed surrogate model, the \textit{Lagrangian Neural Cellular Automaton}. 

\vspace{12pt}

The architecture of our model can be described as a physics-constrained \textit{Graph Neural Cellular Automaton}, a popular kind of generalization of the classical Cellular Automaton. 
\textit{Classical} CAs are discrete dynamical systems generally characterized by a grid of cells, each possessing a state that updates synchronously according to a localized, uniform rule based on the states of the neighboring cells. 
These classical CAs are well-documented substrates for so-called ``emergent'' behavior \cite{Wolfram1983}, and the underlying principle of simulating self-organizing phenomena from simple, localized dynamics has been extended to greater generality and applied in a wide variety of scientific settings \cite{Gilpin2019, Hartl2026}.

In recent years, the parameterization of the local update rule via deep neural networks has given rise to \textit{Neural Cellular Automata} (NCA) \cite{Gilpin2019, Mordvintsev2020}. NCAs have demonstrated remarkable success in learning emergent behaviors, including morphogenesis, texture synthesis, and the emulation of complex fluid dynamics. Because the update rule is strictly local and translationally equivariant, NCAs are highly parallelizable and naturally generalize to grid sizes larger than those seen during training.

\begin{figure}[b]
  \centering
  \includegraphics[width=1.0\linewidth]{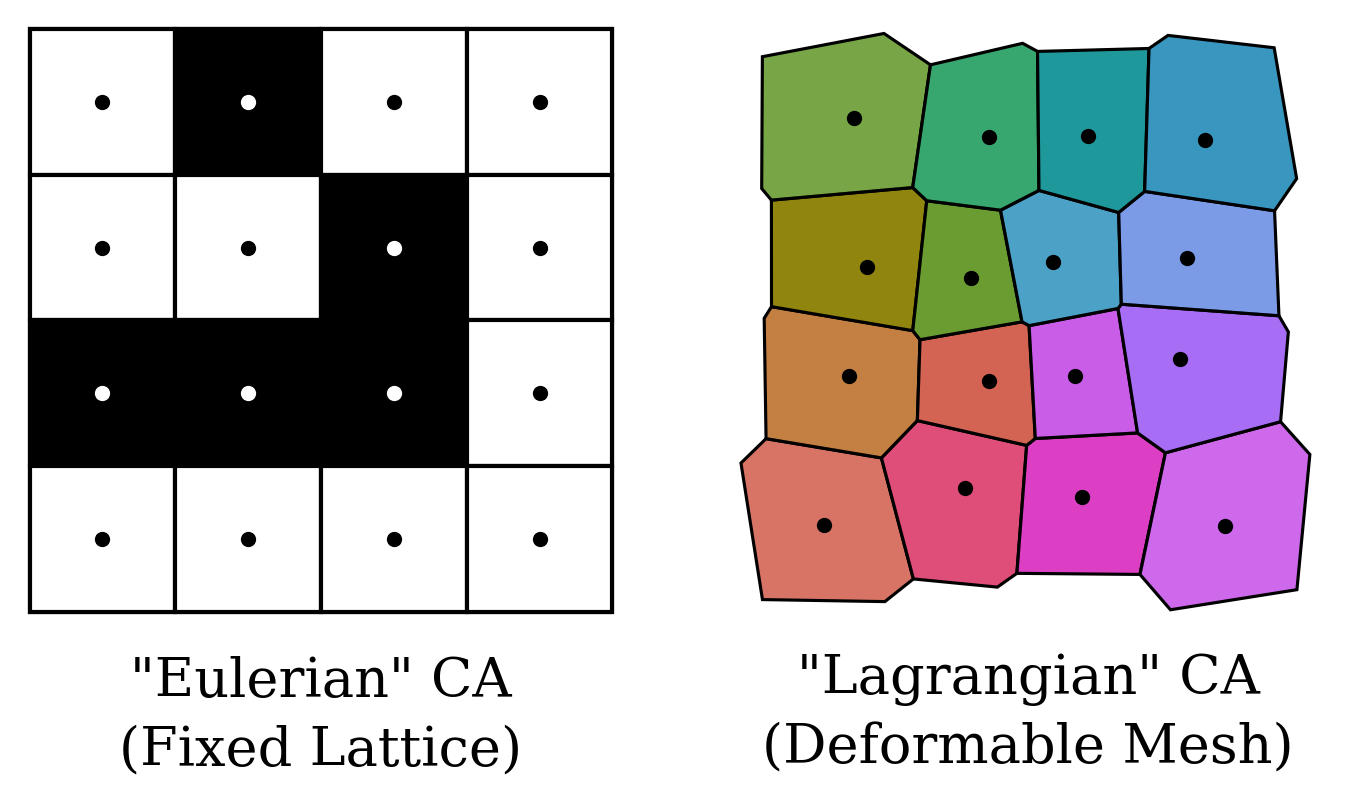}
  \caption{Cartoon representation of the central difference between a traditional ``Eulerian" Cellular Automaton and the ``Lagrangian" Automaton used in this work. In the latter case, we allow the cells to be displaced according to the update rule, while holding neighborhood connectivity fixed. }
  \label{fig:CA_vs_LCA_cartoon}
\end{figure}

However, standard NCAs operate in a static Eulerian reference frame, where the grid represents fixed spatial coordinates and the state vector represents quantities like density or color flowing through the cells. In systems dominated by high-density contrasts and large-scale bulk flows (such as cosmic structure formation) Eulerian NCAs suffer from severe numerical diffusion and advection errors. They also struggle to maintain Galilean invariance, requiring deep, expansive receptive fields to model mass moving across large distances. 

\newpage

To circumvent the limitations of fixed spatial grids, Graph Cellular Automata (GCAs) and related Message Passing Neural Networks extend the CA paradigm even further to arbitrary topologies \cite{Grattarola2021}. 

Graph CAs have been shown to be very successful as surrogate models for fluid simulations and reliable frameworks for modeling phenomena like the flocking of birds where the dynamics of individual particles or agents must be tracked. Their particle structure makes them better suited for and easier to train on point-cloud-like datasets (relative to grid-like CAs). Graph CAs can therefore be naturally trained on particle-based simulation codes and can even be applied as extensions of the particle codes themselves. Because the data structure itself moves with the flow of information, these models are better suited to applications involving large dynamic ranges, like our simulations of cosmic structure. 

However, in a true GCA, nodes can move continuously through space, and connections are often dynamically recomputed based on physical proximity. While dynamically connected GCAs provide an excellent physical inductive bias for particle-based simulations \cite{Sanchez2020}, they introduce extreme computational bottlenecks. The continuous re-evaluation of the graph topology and the reliance on irregular scatter-gather memory operations severely limit their scalability, particularly for cosmological simulations which routinely track $10^9$ to $10^{12}$ particles.

\vspace{12pt}

Our model structure inherits from both of these paradigms and benefits from the computational efficiency of an ``Eulerian'' CA while simultaneously possessing the dynamic expressibility of a Graph CA. As we describe below, our Lagrangian Neural Cellular Automaton allows the nodes to be freely perturbed but holds neighborhood connectivity absolutely fixed, allowing all local updates to be expressed as convolutions on the initial grid.

\begin{figure}[b]
  \centering
  \includegraphics[width=1.0\linewidth]{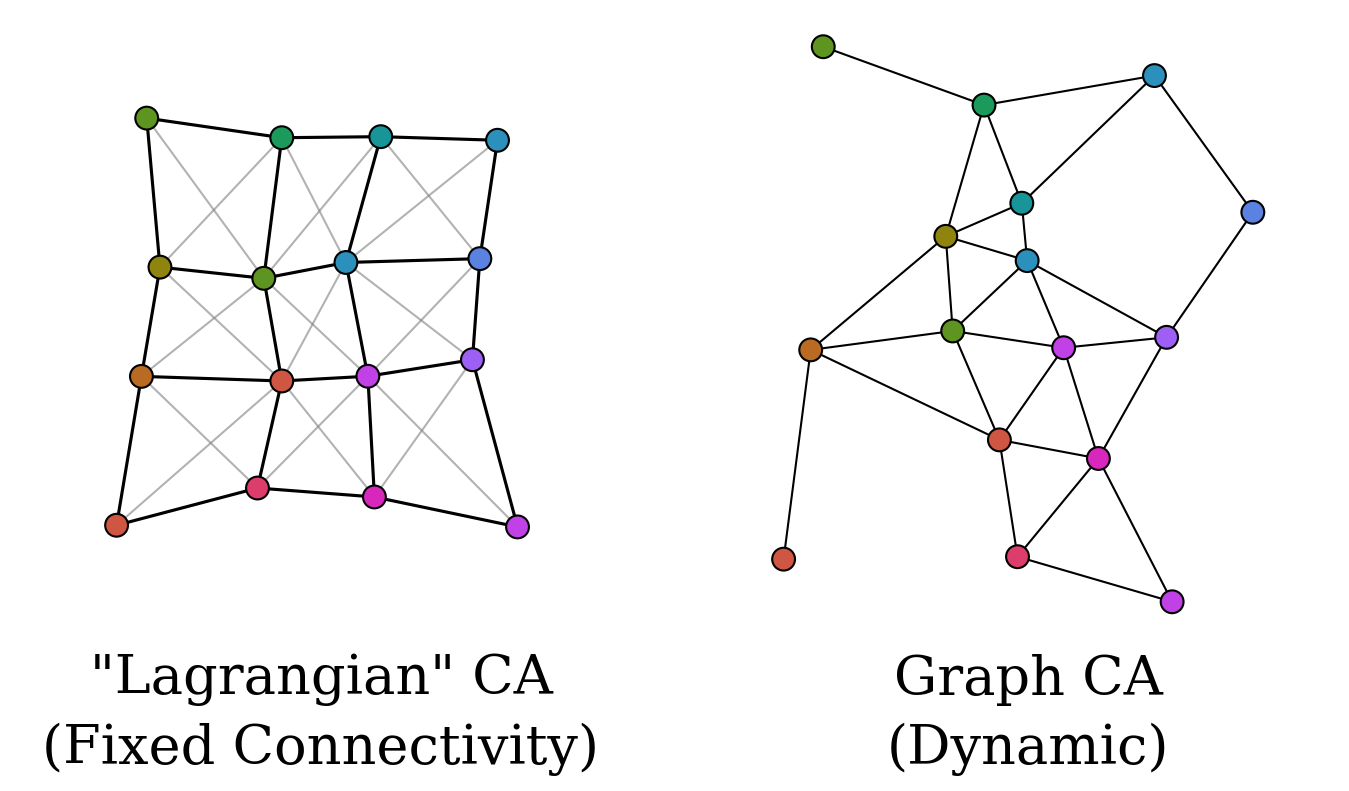}
  \caption{Cartoon representation of the difference between an arbitrary graph connectivity typical of a GCA and the fixed connectivity used here. 
  %In the former case, the edges of the graph are dynamically recalculated based on the relative positions of the nodes (Delaunay triangulation shown here). In the later case, we hold the graph set of connections fixed, even as the positions are updated. 
  In the left panel we show connections between nodes in black if they were initially nearest neighbors, and next-nearest neighbors in grey.
  }
  \label{fig:graph_vs_lattice}
\end{figure}

\begin{figure*}[t]
  \centering
  \includegraphics[width=0.99\linewidth]{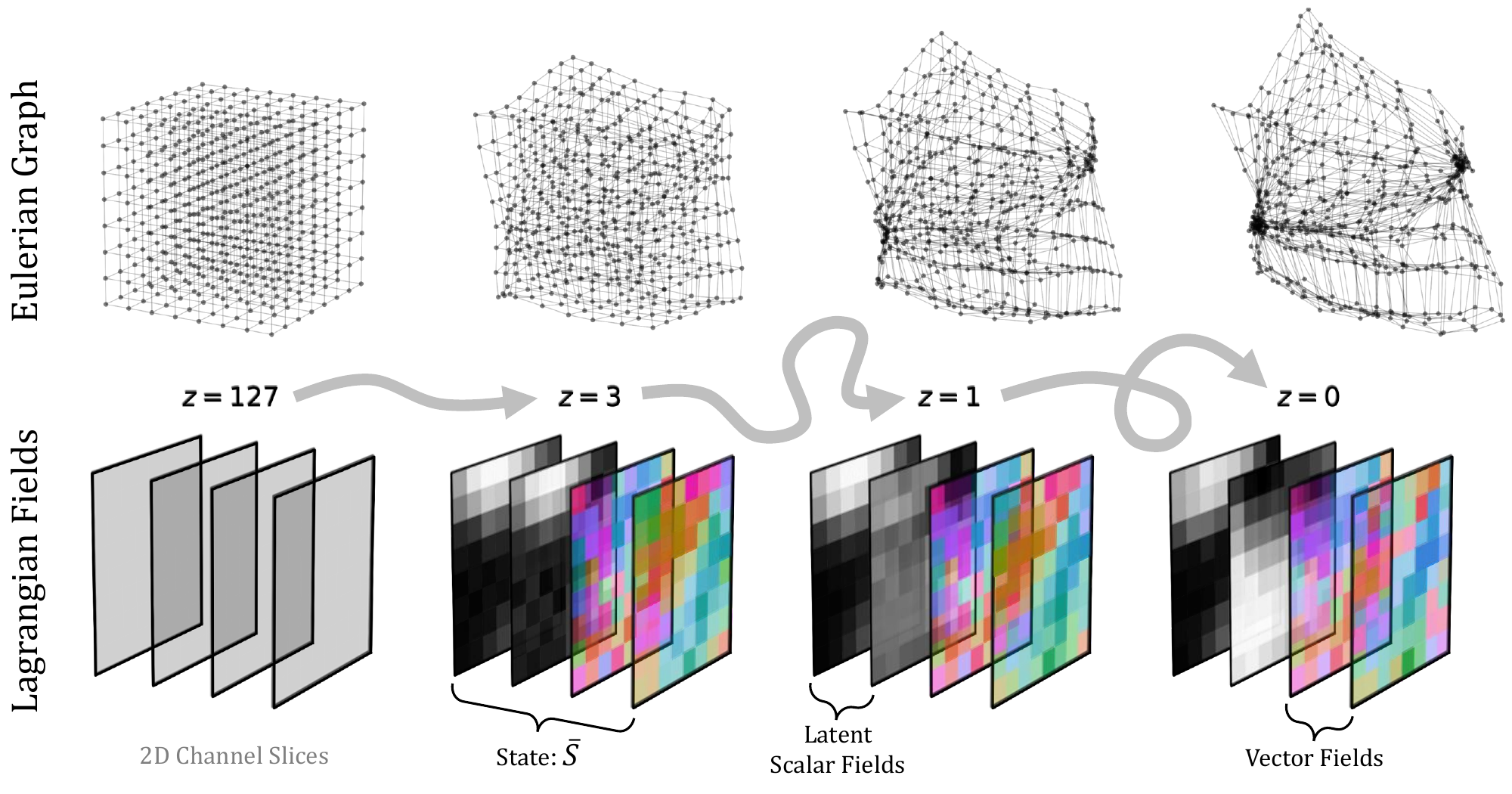}
  \caption{Representations of the dynamic evolution of the LNCA model. The top row shows the evolution of the forward model as a graph of connected nodes in comoving Eulerian coordinates, much like traditional N-body simulation, but with lines connecting initial nearest neighbors. The bottom row visualizes a channel-map of the Lagrangian latent scalar and vector fields along a central slice of the volume. Forward evolution is emulated by the interaction of these fields over many substeps.
  %\oliver{This could be figure 1, it's the main point of the method, and you want people to be intrigued by it}  
  }
  \label{fig:forward_sim}
\end{figure*}
%\newpage

%\newpage
\subsection*{The Lagrangian Neural Cellular Automaton}

\noindent To achieve a scalable, differentiable, physically constrained, and temporally continuous forward model, we introduce the Lagrangian Neural Cellular Automaton (LNCA).  
We describe the unperturbed initial positions of fluid elements by a regular Cartesian grid of Lagrangian coordinates, $q$. In the LNCA, this grid topology is immutable; a node $i$ is permanently connected to its neighbors in $q$-space, regardless of where the node moves in physical Eulerian space, $x$. This fixed topology allows us to replace slow graph operations with highly optimized 3D convolutions, while preserving the continuous particle tracking of a graph-based network.

To suitably capture the complex non-linear physics, the network is endowed with a rich \textit{internal latent feature space}. At any discrete time step $t$, the state of node $i$ is defined by the tuple:
\begin{equation}
\bar{S}_i(t) = \{ \, \Psi_i(t), \, \dot{\Psi}_i(t), \, \mathcal{S}_i(t), \,\mathcal{V}_i(t) \, \}
\end{equation}
where $\Psi_i(t) = x_i(t) - q_i$ is the physical displacement vector and $\dot\Psi_i(t)$ is the comoving velocity.
%\oliver{perhaps write vectors in bold or similar? it's a tad confusing to have $x_i$ not imply the $i$-th coordinate of the 3-vector $x$. Also $v_i = \dot\Psi_i$?} 
The features $\mathcal{S}_i \in \mathbb{R}^{L_s}$ and $\mathcal{V}_i \in \mathbb{R}^{L_v \times 3}$ represent $L_s$ latent scalars and $L_v$ latent vectors, respectively. These latent fields allow the network to accrue and propagate memory of the local deformation history, and to learn internal representations of non-trivial local dynamics which cannot be encoded by the displacement field alone.
%\oliver{Some intuition with the Lagrangian equations of motion might be useful here. There's a second-order ODE for $\psi$ (or two first-order for $\Psi$, $\dot\Psi$) involving the potential that might be useful to quote, to see how and why we need these latents}

To enforce physical consistency on macroscopic scales, the network is tasked only with predicting the \textit{residual} acceleration required to correct the Zeldovich approximation (ZA), rather than learning the total gravitational dynamics from scratch. The total Lagrangian displacement field is decomposed as:
\begin{equation}
\boldsymbol{\Psi}(t) = D(t) \boldsymbol{\Psi}_{\text{IC}} + \boldsymbol{\Psi}_{\text{res}}(t)
\end{equation}
where $D(t)$ is the linear growth factor and $\boldsymbol{\Psi}_{\text{IC}}$ is the initial linear displacement field.

\vspace{12pt}

Furthermore, to ensure temporal continuity, we task the model with predicting the time derivatives of the displacement and latent fields, rather than predicting the fields directly. This allows us to \textit{integrate} the model outputs over many timesteps to produce a realization of the displacement fields with reliable local velocities and which can track the trajectories of the particles over the rollout. %This novel integration-based construction also allows us to naturally account for continuous change in the point-wise timestamp, which is necessary to produce ``lightcones'' which match the nature of real survey catalogs wherein nearby structures appear to us more developed than those billions of light years further away.
The state is advanced by a differentiable symplectic update over discrete steps $\Delta t$:
\begin{align}
    \dot{\Psi}_{i,\text{res}}(t + \Delta t) &=  \dot{\Psi}_{i,\text{res}}(t) + \ddot{\Psi}_{i,\text{res}}(t) \Delta t \notag \\
    \Psi_{i,\text{res}}(t + \Delta t) &= \Psi_{i,\text{res}}(t) + \dot{\Psi}_{i,\text{res}}(t + \Delta t) \Delta t \notag \\
    \mathcal{S}_i(t + \Delta t) &= \mathcal{S}_i(t) + \dot{\mathcal{S}}_i(t) \Delta t \\
    \mathcal{V}_i(t + \Delta t) &= \mathcal{V}_i(t) + \dot{\mathcal{V}}_i(t) \Delta t \notag
\end{align}

where $\ddot\Psi, \dot {\cal S}, \dot{\cal V}$ are the pieces predicted by our model.

%By hard-coding the linear displacement history, the long-range dynamics are reproduced exactly which allows our model to be completely localized, as the nearest-neighbor only interactions are not required to support long-range gravity and must only produce local corrective forces to sharpen features and halt the artificial stream-crossing inherent to linear theory. 

\newpage

\subsection*{Equivariant Formulation}

The evolution of the system is governed by a strictly local update function which iteratively evolves the set of displacement and latent fields via an interaction function parameterized by a neural network.  
To guarantee that the model's physical predictions are strictly $E(3)$ translationally and rotationally-equivariant, the network does not operate on the raw state vectors themselves. Instead, we compute a comprehensive basis of geometric invariants and express our learned output in terms of these.

\begin{figure}[b]
  \centering
  \includegraphics[width=1.0\linewidth]{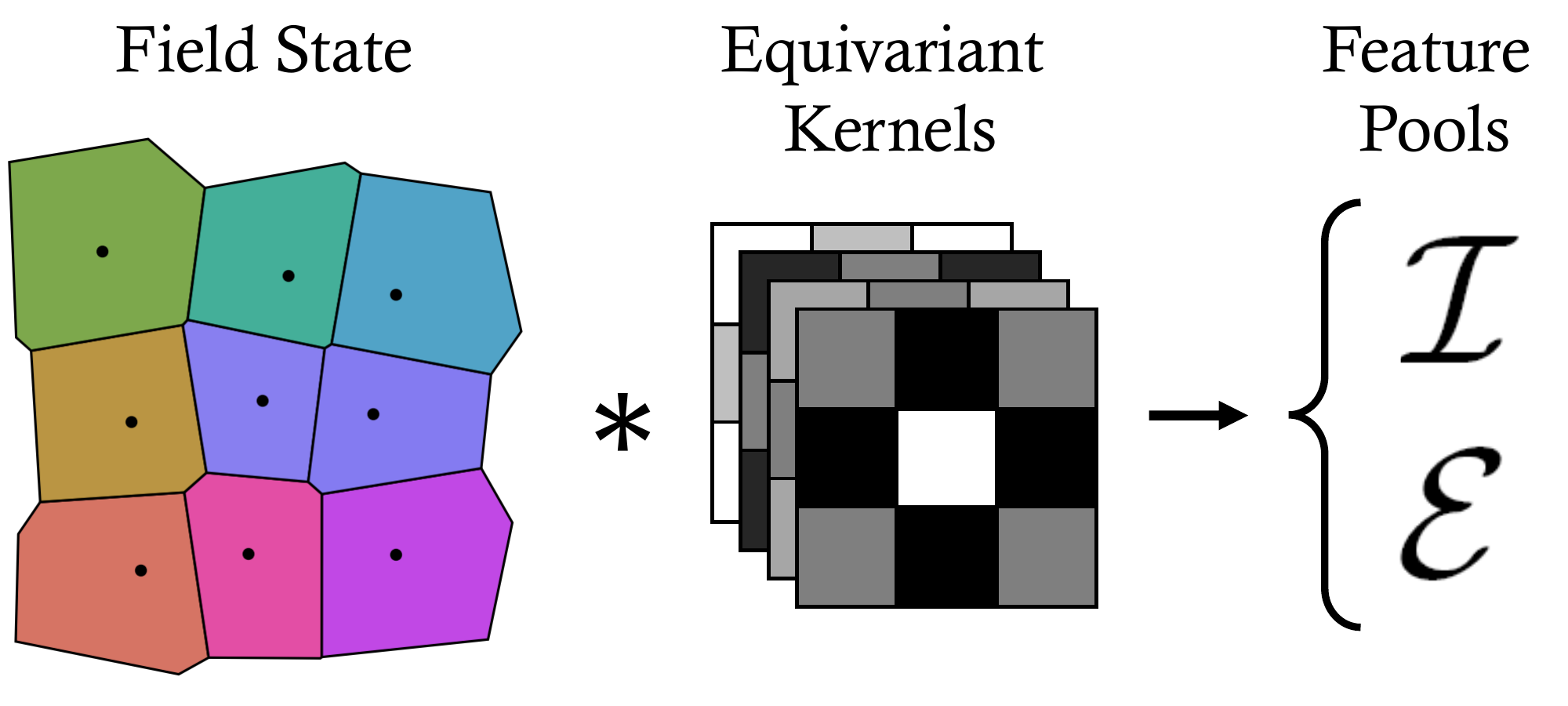}
  \caption{Cartoon diagram of the extraction of equivariant feature pools by convolution with pre-defined  kernels.
  %\oliver{This figure is cute but isn't as instructive to me as the other ones. Not sure if it's needed.}
  }
  \label{fig:equi_conv}
\end{figure}

To construct this basis, we evaluate the set of Laplacians ($\nabla_q^2$), Hessians ($\mathbf{H}_q$), and vector products over the $q$-lattice with fixed finite-difference operators to extract the local kinematics. 
From these local kinematic fields and the latent states, we construct invariant and equivariant feature pools at every node $i$:

\begin{enumerate}
    \item \textbf{A Scalar Pool ($\mathcal{I}_i$):} A comprehensive set of $E(3)$-invariant scalars constructed from gradients and vector dot products, as well as externally-provided global cosmological parameters. 
    \item \textbf{A Vector Menu ($\mathcal{E}_i$):} A basis of equivariant, dimensionally consistent vectors, including the displacement, velocity and Laplacians and vector products of the latent fields. 
\end{enumerate}

\begin{figure}[t]
  \centering
  \includegraphics[width=1.0\linewidth]{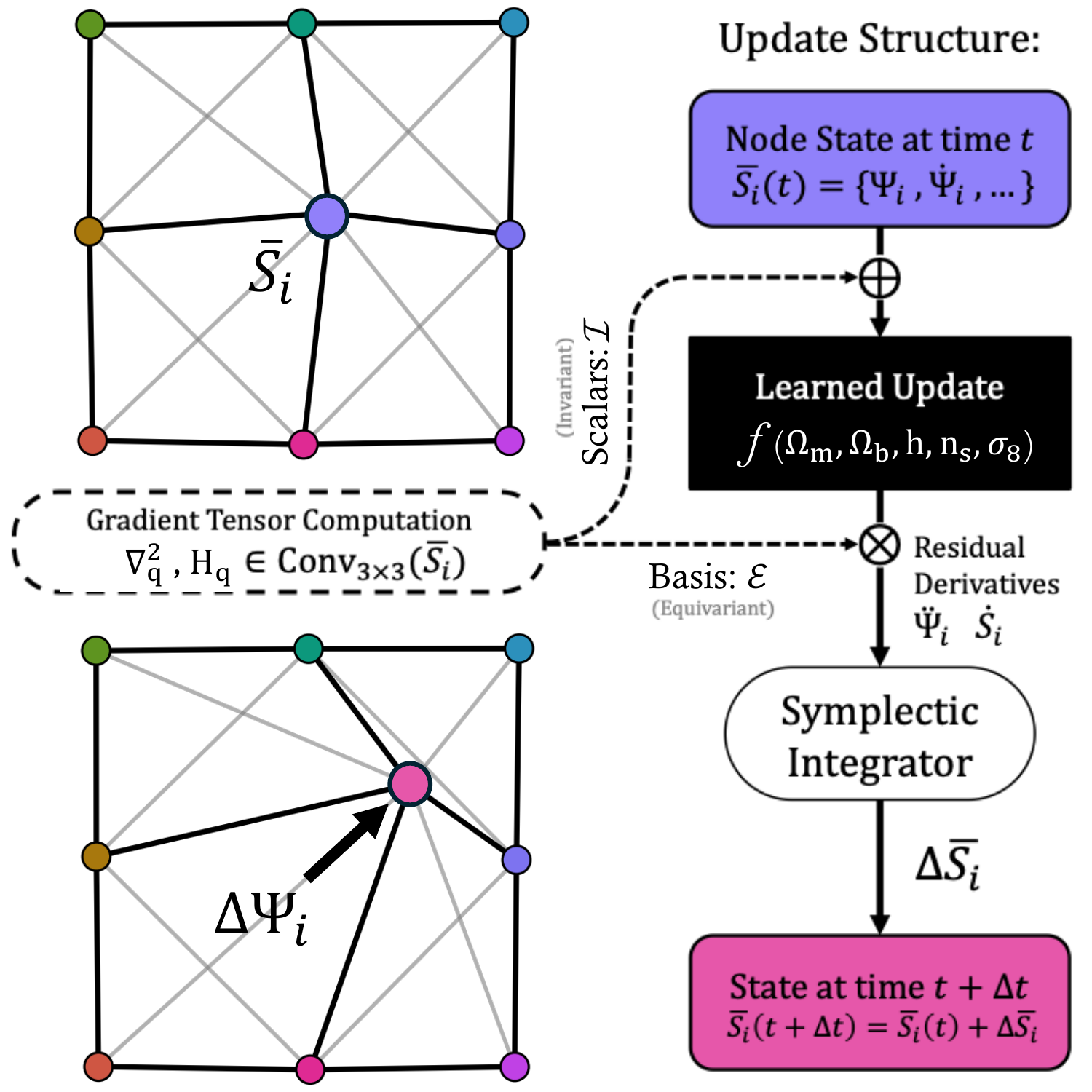}
  \caption{Flow-chart diagram of the update procedure for a single node over a single timestep, detailing the extraction of the equivariant basis and the linear synthesis of the output as derivatives of the latent fields, which are integrated over timestep $\Delta t$ to yield the next state.
  %\oliver{Also I'm a bit unclear of what the + and x operators do; is this transforming into and outof the invariant basis? I would suggest being more clear in this figure about where all the relevant $S_i$, $\mathcal{S},\mathcal{V}$ and other things enter, as it's pretty nuanced.}
  }
  \label{fig:update_rule}
\end{figure}

The feature pools are constructed exclusively from first and higher-order derivatives of the displacement field $\Psi$, not $\Psi$ itself. Since at linear order the displacement is the gradient of the gravitational potential, $\Psi \sim \partial \phi$, the minimum feature order is $\partial \Psi \sim \partial^2 \phi$ (the tidal field), and the model is therefore constrained by the equivalence principle by construction.

The scalar pool contains the invariants that source Lagrangian perturbation theory at second and third order, including $\mathrm{Tr}(s^2)$, $\text{Tr}(K^2)$, $\text{Tr}(K^3)$, $\mathrm{det(s)}$, and the velocity field analogs, ensuring the model can reproduce third-order Lagrangian perturbation theory while being prohibited by construction from coupling to any long-range modes.

\newpage

The network receives these scalar invariants $\mathcal{I}_i$ as input and maps them to a set of output coefficients:

\begin{equation}
\{ \dot{\mathcal{S}}_i \, , \mathbf{W}_i \}= f_\theta(\mathcal{I}_i(t))
\end{equation}

\noindent
where $f_\theta$ is a high-dimensional function approximated by the deep learning model. All learned behavior of the model is restricted to this point-wise scalar function $f_\theta$, to guarantee symmetry.

Once the output weight coefficients $\mathbf{W}_i$ have been computed, the physical displacement and latent vector derivative updates are then synthesized via linear combinations of the elements from the vector menu:
\begin{equation}
\begin{split}
    \{ \ddot{\boldsymbol{\Psi}}_{i,\text{res}}, \dot{\mathcal{V}}_i \}&= \sum_{k} \mathbf{W}_{i,k} \mathcal{E}_{i,k} 
\end{split}
\end{equation}

By operating strictly point-wise on this rich rotationally invariant scalar pool $\mathcal{I}_i$, the model is completely agnostic to the global coordinate frame and purely translation invariant by construction. 
Because the scalar coefficients are perfectly invariant, and the menus are composed of strictly equivariant vectors, the resulting synthesized updates are mathematically guaranteed to be $E(3)$ rotationally equivariant.

%\oliver{I think it would be useful to be a bit more clear about what quantity is used for what. e.g., from the current text, it's not super clear that the real pipeline is (I think): state -> $I_i$ [which depends only on $\Psi,\dot\Psi$ or also on the vectors? I got a bit lost here] -> updates for scalar-latent -> updates for vector-latent given vector coeffs (which I think only appear here?) -> updates for $\ddot\Psi$. Also you don't explicitly have $\dot S$ anywhere, even though it appears in the figure (which you should link to in the text)}

%\oliver{The \textit{exact} evolution equation for $\Psi$ is given by $\Psi''(\vec q, \tau) + \mathcal{H}(\tau)\Psi'(\vec q,\tau) = -\nabla_{\vec x}\phi(\vec x,\tau)$ (in conformal time), where $\vec x = \vec q + \Psi(\vec q, \tau)$. It would be a good exercise to check that this form is consistent with the symmetry structure / assertions that you're making above, and that you can recover, e.g., the linear-order solution from the variables within $\mathcal{S}$ and $\mathcal{I}$ (indeed, it would be useful to say what these variables are)}

%\newpage

%Because the integration is temporally continuous, the LNCA naturally supports the generation of past lightcones for direct comparison with observational survey data.

\newpage

\subsection*{Update Network Structure}
\noindent The local update rule $f_{\theta}$ at the heart of the LNCA is a compact equivariant network
that maps the local geometric state of each Lagrangian node to a set of scalar and vector field updates.
As mentioned above, the learnable function $f_{\theta}$ operates strictly on a large pool of
feature scalars, $\mathcal{I}_i$.
Its design is motivated by a desire for both strict analytic continuity and interpretable sparsity, ensuring that the learned dynamics are physically constrained and can be read as an
explicit algebraic expression over physically meaningful invariants. To achieve this, we employ a constrained version of a Kolmogorov-Arnold Network (KAN) \cite{Liu2025, Liu2025b} to encode the learnable function $f_{\theta}$, in place of a more standard Multi-Layer Perceptron (MLP). In this formulation, each network layer replaces the standard (ReLU) activation function with a learned polynomial, $\sigma$:

\begin{equation}
  \sigma(\boldsymbol{x}) =
    \sum_{p=1}^{p_{\rm max}} \mathcal{C}^{(p)} \cdot \frac{\boldsymbol{x}^p}{p!}
    + \boldsymbol{b}.
  \label{eq:taylorkan}
\end{equation}

The polynomial structure of each layer means that every output neuron is, in principle, an explicit Taylor expansion of the input invariants, making the learned update rule directly inspectable. This allows the network to capture non-linear behavior much more exactly and with far fewer learned parameters.
The network first compresses the scalar pool to a bottleneck of width $16$, expands to a hidden representation of width $32$, and projects to a vector of output coefficients, $\{\dot{\mathcal{S}}_i, \mathbf{W}_i \}$.
Layer normalization is applied after both the input and bottleneck projections. A cartoon of the model structure is shown below, highlighting the polynomial learned activation functions.

\begin{figure}[b]
  \centering
  \includegraphics[width=\columnwidth]{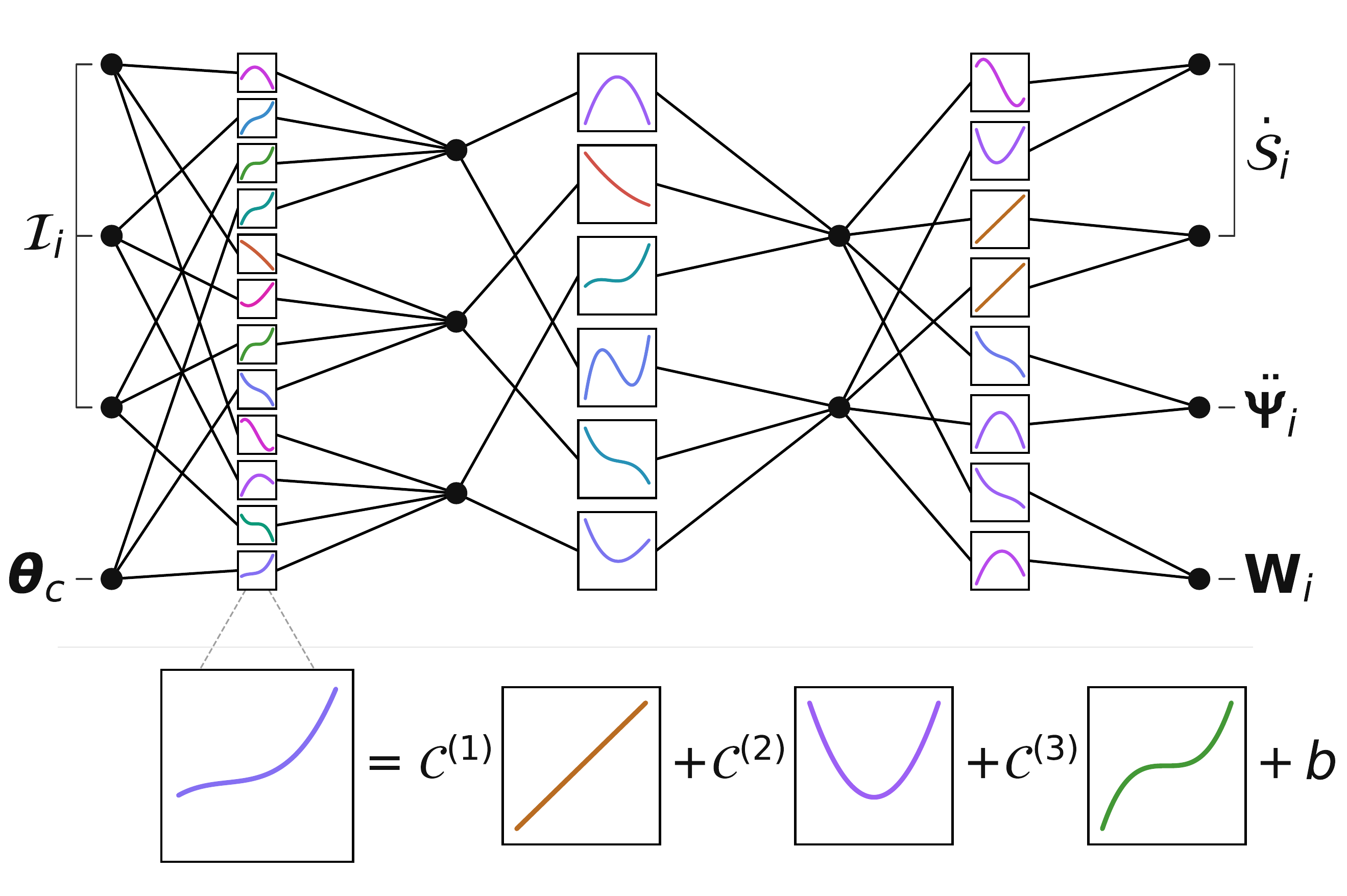}
  \caption{
    KAN architecture of the LNCA update network $f_{\theta}$. The invariant scalar pool $\mathcal{I}_i$ and cosmological parameter vector $\boldsymbol{\theta}_c$ are passed through three TaylorKAN layers
    whose output is decomposed into scalar weights that linearly modulate the equivariant basis to produce the
    field updates.
  }
  \label{fig:network_architecture}
\end{figure}

%\newpage

\subsection*{Training Data and Cosmological Conditioning}

\noindent To optimize the parameters of the Lagrangian Neural Cellular Automaton (LNCA), we employ a traditional
supervised learning method, utilizing the Quijote simulation suite \cite{Villaescusa_Navarro_2020}
as ground truth.  Specifically, we leverage the Latin Hypercube set, which spans an extremely broad prior volume over the standard five-parameter $\Lambda$CDM cosmology:
$\boldsymbol{\theta}_c = \{\Omega_m, \Omega_b, h, n_s, \sigma_8\}$.  By training a single,
unified network across this diverse distribution of cosmologies, the LNCA learns a generalized
physical mapping rather than overfitting to a specific expansion history.

Global cosmological dependence is injected directly into the local update rule.  For any given simulation volume, the parameters $\boldsymbol{\theta}_c$ are standardized and globally broadcast
to every node in the Lagrangian grid, where they are appended to the rotationally invariant scalar
pool $\mathcal{I}_i$ available to the update network.  Consequently, the network learns to
dynamically modulate the amplitudes and relationships between its structural corrections based on these macroscopic properties.

%\subsection*{Forward Model and Initial Conditions}

The forward model is formulated as an initial value problem.  The network is initialized at an
early scale factor $a_0 = 1/128$ using the initial linear displacement field $\boldsymbol{\Psi}_{\rm IC}$
used by the corresponding Quijote simulation. The LNCA integrates forward in
continuous scale-factor steps to $a = 1.0$, producing continuous particle trajectories through the
coupled update:
\begin{equation}
  f_{\theta}\!\left(\boldsymbol{\Psi}_{\rm tot},
      \dot{\boldsymbol{\Psi}}_{\rm res}, \mathcal{S}, \mathcal{V};\,
      \boldsymbol{\theta}_c\right) 
      \, \rightarrow \,
      \ddot{\boldsymbol{\Psi}}_{\rm res} 
      \, \rightarrow \,
  \dot{\boldsymbol{\Psi}}_{\rm res}  =  \int\ddot{\boldsymbol{\Psi}}_{\rm res}dt
  \label{eq:lnca_ode}
\end{equation}

%where $\boldsymbol{\Psi}_{\rm tot} = D(t)\boldsymbol{\Psi}_{\rm ZA} + \boldsymbol{\Psi}_{\rm res}$ is the total displacement, $D(t)$ is the linear growth factor computed numerically for each cosmology in the batch, and $\mathcal{S}$ and $\mathcal{V}$ denote the latent scalar and vector fields that carry memory of the preceding dynamical state.

To compute the loss, predictions are extracted at five snapshot redshifts
($z \in \{3.0, 2.0, 1.0, 0.5, 0.0\}$) and compared directly against the corresponding non-linear
$N$-body outputs.  Because the state at redshift $z$ depends recursively on all preceding substeps,
the network is trained end-to-end via Backpropagation Through Time (BPTT) through the full
integration trajectory.

The model is trained on batches of sub-volume crops of the Quijote simulations of width $200 \,\text{Mpc}/h$.
Because these crops are non-periodic, a critical concern is the contamination of boundary nodes by finite-difference artifacts.  Because the finite difference stencil kernels
(gradients, Laplacians, Hessians) possess a finite spatial support, unrolling the recurrence over $N_{\rm steps}$ integration steps allows boundary errors to propagate inward at a known and finite rate (see further discussion of this feature in Appendix \ref{app:interp}).  To prevent these artifacts from corrupting the loss gradients, we employ a spatial padding strategy: the LNCA evaluates the dynamics on a padded volume of $96^3$ voxels, but strictly crops boundary margin of $N_{\rm steps}$ voxels on each face before evaluating any loss term, restricting all gradient signals to the uncontaminated central core.

\subsection*{Multi-Objective Loss Function}

\noindent The training curriculum of a Physics-informed ML model is just as important as the design of the model itself. While we can guarantee certain properties like rotational equivariance from the model design, the choice of loss function allows us to further specify which properties of the target are most important and thus specifically incentivize how our model should behave. 

In our context, there are several complementary features of the output that we would like to optimize for simultaneously. 
We train our model against the target N-body simulations, compared relative to the Zeldovich predictions in four objectives: per-particle displacement and momentum, real-space density, and halo morphology. Each component is normalized relative to the Zeldovich Approximation (ZA) such that $\mathcal{L}_{\{X\}} = 1$ indicates parity with ZA at fixed cost and $\mathcal{L}_{\{X\}} < 1$ indicates genuine improvement.

\subsubsection*{Per-Particle Displacement Loss $(\mathcal{L}_{\rm \Psi})$}

\noindent The primary objective penalizes the residual difference between the predicted $\boldsymbol{\Psi}_{\rm pred}$ and target $\boldsymbol{\Psi}_{\rm true}$ \emph{non-linear}
displacement, normalized by the input Zeldovich approximation (ZA) baseline $\boldsymbol{\Psi}_{\rm ZA}$ so that the loss is dimensionless and cosmology-independent in scale:
\begin{equation}
  \mathcal{L}_{\rm \Psi}
    = \frac{
        \bigl\langle \bigl\|
          \boldsymbol{\Psi}_{\rm pred} - \boldsymbol{\Psi}_{\rm true}
        \bigr\|^2 \bigr\rangle_{\mathbf{q}}
      }{
        \bigl\langle \bigl\|
          \boldsymbol{\Psi}_{\rm ZA} \; - \boldsymbol{\Psi}_{\rm true}
        \bigr\|^2 \bigr\rangle_{\mathbf{q}}
      }
  \label{eq:loss_disp}
\end{equation}
where $\boldsymbol{\Psi}_{\rm ZA} = \boldsymbol{\Psi}_{\rm IC}\,D(t)$ is the linear approximation and $\boldsymbol{\Psi}_{\rm pred} = \boldsymbol{\Psi}_{\rm ZA} + \boldsymbol{\Psi}_{\rm LNCA}$ is the combined prediction of our model.
Averages $\langle \, \rangle_{\mathbf{q}}$ are taken over all particles in the core volume and all snapshots in the batch.

\subsubsection*{Per-Particle Momentum Loss $(\mathcal{L}_{\mathbf p})$}

\noindent To better encourage the trajectories to align with the dynamics of the target, we additionally apply a penalty to the per-particle momentum residual, which can be extracted directly from our model's output. As with the above displacement loss, we normalize this term relative to the momentum residual of the linear Zeldovich approximation:
\begin{equation}
  \mathcal{L}_{\mathbf p}
    = \frac{
        \bigl\langle \bigl\|
          \mathbf{p}_{\rm pred} - \mathbf{p}_{\rm true}
        \bigr\|^2 \bigr\rangle_{\mathbf{q}}
      }{
        \bigl\langle \bigl\|
          \mathbf{p}_{\rm ZA} \; - \mathbf{p}_{\rm true}
        \bigr\|^2 \bigr\rangle_{\mathbf{q}}
      }
  \label{eq:loss_mom}
\end{equation}
where $\mathbf{p}_{\rm true}$ is the target N-body momentum, $\mathbf{p}_{\rm ZA}$ is the linear approximation and $\mathbf{p}_{\rm pred} = \mathbf{p}_{\rm ZA} + \mathbf{p}_{\rm LNCA}$ is the combined LNCA momentum prediction. As before, averages are taken over all particles in the core volume and all snapshots in the batch.

\subsubsection*{Real-Space Density Loss $(\mathcal{L}_{ \delta})$}
\noindent For each redshift snapshot in the training data, the predicted Lagrangian displacements are projected on to Eulerian
overdensity fields $\delta(\mathbf{x})$ through differentiable Cloud-in-Cell (CIC) deposition on a uniform grid. These density maps are compared $\delta_{\rm true}$ via:

\begin{equation}
\mathcal{L}_{\delta} \;=\; \frac{\bigl\langle\, (\delta_{\rm pred} - \delta_{\rm true})^2\,\bigr\rangle_{\mathbf{x}}}
                          {\bigl\langle\, (\delta_{\rm ZA}\;\, - \delta_{\rm true})^2\,\bigr\rangle_{\mathbf{x}}}
\label{eq:za_norm}
\end{equation}

Unlike $\mathcal{L}_\Psi$, this term couples particles in the predicted and target fields through their relative final positions, not initial positions, and is thus sensitive to the arrangement of neighboring streams and penalizes the shell-crossing artifacts ZA produces wherever trajectories have collided. Because the CIC deposition is spatially
local, the net gradient contribution of $\mathcal{L}_\delta$ vanishes in uniform voids and sharpens specifically in the filaments and halos where the per-particle displacement becomes non-linear.

\subsubsection*{Halo Population Loss ($\mathcal{L}_{\rm Halo}$)}

To directly reward the collapse of individual bound structures, we define a new term applied only to particles identified as belonging to Halos in the target field. For each halo
$h$ we form four moments of this set: the center-of-mass position $\bar{\mathbf{x}}_h$, the
bulk momentum $\bar{\mathbf{p}}_h$, and the root-mean-square dispersions $r_{x,h}$ and
$\sigma_{p,h}$ of the member positions and momenta about those means. Each is compared to its
target value and ZA-normalized as before:
\begin{equation}
\mathcal{L}_{\rm Halo} \;=\; \frac{1}{4}\!\!
  \sum_{\mu \,\in\, \{\bar{\mathbf{x}},\,\bar{\mathbf{p}},\,r_x,\,\sigma_p\}}\!\!
  \frac{\bigl\langle\,\lvert \mu_{\rm pred} - \mu_{\rm true}\rvert^{2}\,\bigr\rangle_{h}}
       {\bigl\langle\,\lvert \mu_{\rm ZA}\; - \mu_{\rm true}\rvert^{2}\,\bigr\rangle_{h}}
\label{eq:lhalo}
\end{equation}
where $\langle\,\rangle_h$ averages over all target halos across snapshots in the batch,
$\lvert\cdot\rvert$ is the Euclidean norm for the vector moments and the absolute value for
the scalar dispersions, and the $1/4$ preserves $\mathcal{L}_{\rm Halo}=1$ at ZA parity.

Depending only on these low-order moments of a fixed set, the term is invariant to any permutation of a halo's particles. This means the model is rewarded for gathering the correct mass into a compact ($r_x$) dynamically hot ($\sigma_p$) clump at the right location ($\bar{\mathbf{x}}$) and bulk velocity ($\bar{\mathbf{p}}$), without requiring that each particle lands at exactly its corresponding location in the target.

%This is the coarse phase-space signature of a virialized object, and it separates whether a bound halo has formed from the far stricter question, already penalized by $\mathcal{L}_\Psi$ and $\mathcal{L}_{\mathbf{p}}$, of whether every particle is placed exactly. 

%That interior is intrinsically hard to reproduce: once member orbits undergo shell-crossing they are chaotic, so the conditional expectation of any single final coordinate given the coarse inputs is nearly null. A deterministic $L_2$ fit therefore regresses toward the smooth conditional mean and systematically under-produces the internal scatter, leaving halos too diffuse and kinematically cold. Matching $s_x$ and $s_p$ makes that dispersion an explicit, permutation-invariant target rather than an unwinnable per-particle penalty.

\vspace{12pt}

These four terms are combined to yield our total, physics-informed loss function. The LNCA model is trained using a hybrid evolutionary scheme to balance exploitation of successful latent fields with exploration of possible functions. The full evolution method, optimizer schedule, regularization,
and curriculum strategies are detailed in Appendix~\ref{app:training}.

\begin{figure*}[thp]
  \centering
  \includegraphics[width=\textwidth]{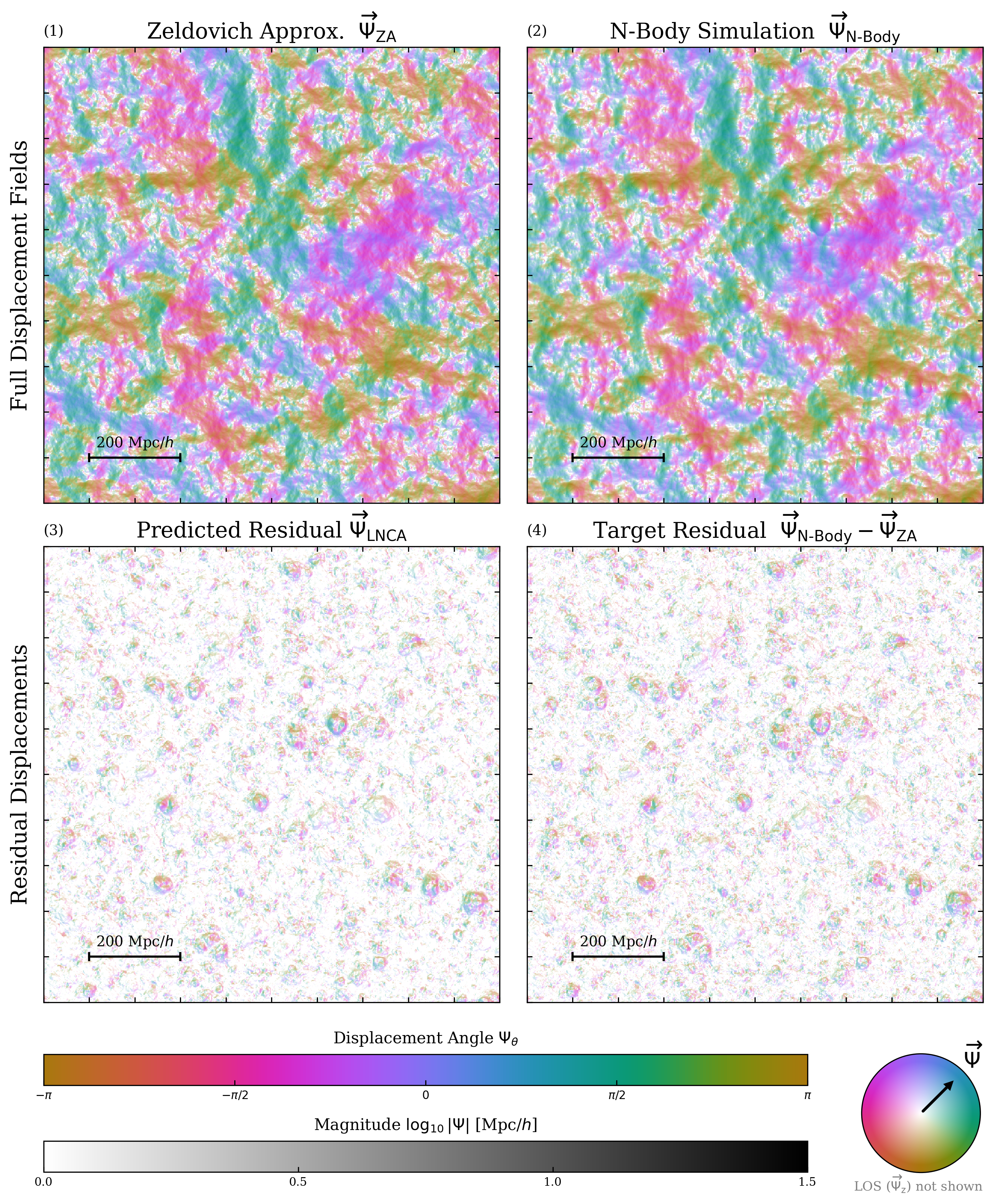}
  \caption{Visualization of the Lagrangian displacement fields for a  linear ``Zeldovich" displacement (1st Panel) and a full-gravity N-body simulation (2nd Panel) using the same initial conditions. The 4th Panel shows the difference between these two displacement fields, which is the target of our model. The 3rd panel shows the displacement field predicted by our Neural Automaton. We visualize displacement vectors here by coloring the Lagrangian coordinate ($\boldsymbol{q}$) by the angular phase (hue) and magnitude (saturation/opacity) of the local displacement vector $\boldsymbol{\Psi}(\boldsymbol{q})$ or residual.}
  \label{fig:predicted_vs_target_residuals}
\end{figure*}

\section{Validation Results}\label{sec:validation}

\noindent We assess our trained model's performance with the following metrics, situated relative to the input ZA and N-body truth. Each comparison probes a distinct facet of the learned map from initial conditions to final evolved particle configurations. The displacement residual diagnoses how well the network reconstructs small-scale non-linear features (\textit{i.e.}\ shell crossings) on top of the Zeldovich background. The matter power spectrum $P_\delta(k)$ characterizes amplitude recovery across scales and redshifts, while the cross-correlation coefficient $r(k)$ measures the phase agreement between predicted and reference density fields independently of any amplitude offset. %Throughout, we benchmark against ZA as a linear-order floor and relative to the corresponding Quijote Latin Hypercube realizations which are held out from training.

\subsection*{Displacement Residuals}\label{sec:validation:displacements}

\noindent Figure~\ref{fig:predicted_vs_target_residuals} compares our LNCA model's output non-linear displacement correction $\bm{\Psi}_{\mathrm{LNCA}}(\bm{q})$ to the ground-truth residual $\bm{\Psi}_{\text{N-body}}(\bm{q})-\bm{\Psi}_{\mathrm{ZA}}(\bm{q})$, where $\bm{q}$ denotes the Lagrangian coordinate of each particle. By construction, both fields vanish in the linear regime, so any structure visible in the lower panels reflects genuinely post-Zeldovich dynamics. The most striking feature is the network's ability to reconstruct the bubble-like shells that correspond to regions of  shell crossing. These caustic surfaces, where neighboring streams in the cold dark matter sheet fold onto one another, are precisely the structures that linear Lagrangian theory cannot represent. The accurate placement and amplitude of these features are a strong indication that our model has learned a meaningful  non-linear correction rather than a smoothed-out averaging of the target field. Filament walls, void boundaries, and the curved sheets surrounding collapsing regions are all recovered with sub-cell positional accuracy and with residual amplitudes that closely track the N-body reference.

On larger scales, the figure shows the coherent infall pattern around overdense regions and the corresponding correction of the surrounding void with high fidelity. This regime is dominated by mildly non-linear evolution, where higher-order contributions become important; the close agreement here is consistent with the network having implicitly captured information equivalent to several perturbative orders, without recourse to an explicit expansion. We observe no systematic bias toward over- or under-prediction of the residual magnitude across the range of environments sampled in the test set, and the orientation of the displacement vectors is preserved even in regions of high local shear. The largest errors are systematically found within halos, where the exact positions of particles are most difficult to predict due to the extremely non-linear sub-halo dynamics and where specific particle identities can be safely mixed without impacting the downstream density field.

\newpage

\begin{figure}[t]
  \centering
  \includegraphics[width=1.0\linewidth]{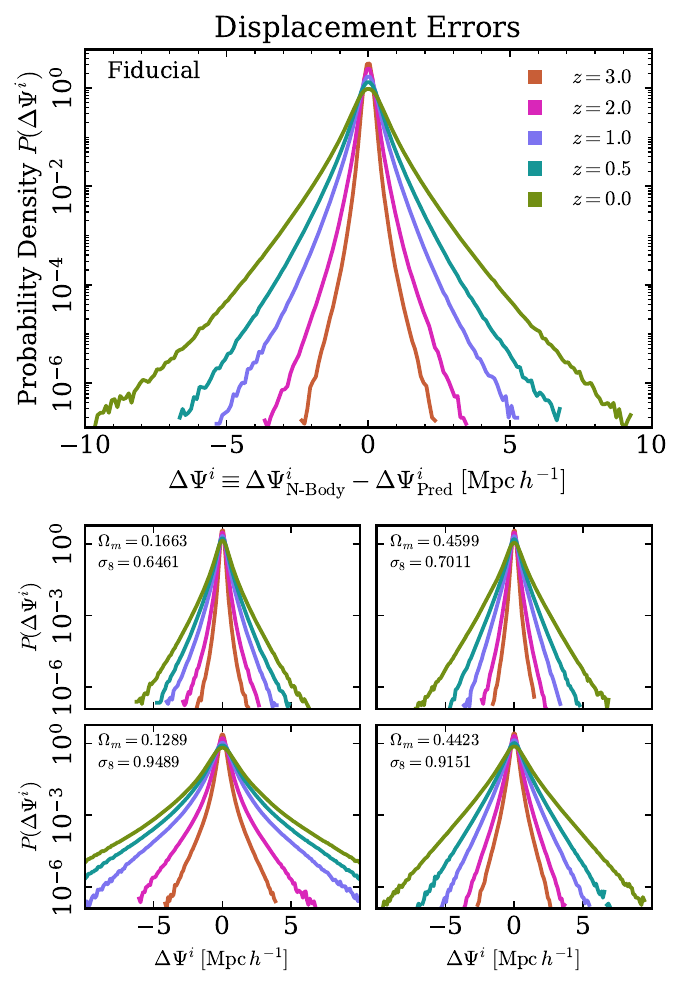}
  \caption{Probability density functions for the component-wise displacement errors between the predicted $\Psi_{\text{Pred}} = \Psi_{\text{ZA}} + \Psi_{\text{LNCA}}$ and target $\Psi_{\text{N-body}}$ displacement vector fields. The large top panel highlights our results on fiducial cosmology while the lower panels display our results on held-out cosmologies, extremely far from the fiducial baseline.} 
  \label{fig:delta_disp}
\end{figure}

To further quantify the agreement between our prediction and the target N-body simulation at the level of the displacement fields, we take the particle-wise difference between the target and output vector field components, $\Delta \Psi^i \equiv \Psi_{\text{N-body}}^i - \Psi_{\text{Pred}}^i$. For fiducial cosmology, the median displacement error is a modest $|\Delta \Psi^i| \simeq 0.2 ~\text{Mpc}/h$. We plot the distributions of these particle position errors in Figure \ref{fig:delta_disp} above, as a function of redshift.
In all cases, we note that the absolute difference of a typical ($95\%$) particle is $|\Delta \Psi^i| < 2 \, \mathrm{Mpc} \, h^{-1}$, with only $5\%$ of particles being misplaced by any more. For reference, the RMS residual displacements of a fiducial target is $|\Psi_{\text{N-body}} - \Psi_{\text{ZA}}| \sim 2 \, \mathrm{Mpc} \, h^{-1}$.
We note, as expected, that the width of these error distributions around zero behaves systematically as a function of redshift, where snapshots closer to $z=0$ are subject to greater degrees of non-linear structure formation and are thus more difficult for the model to exactly reproduce.

\begin{figure*}[thp]
  \centering
  \includegraphics[width=0.99\linewidth]{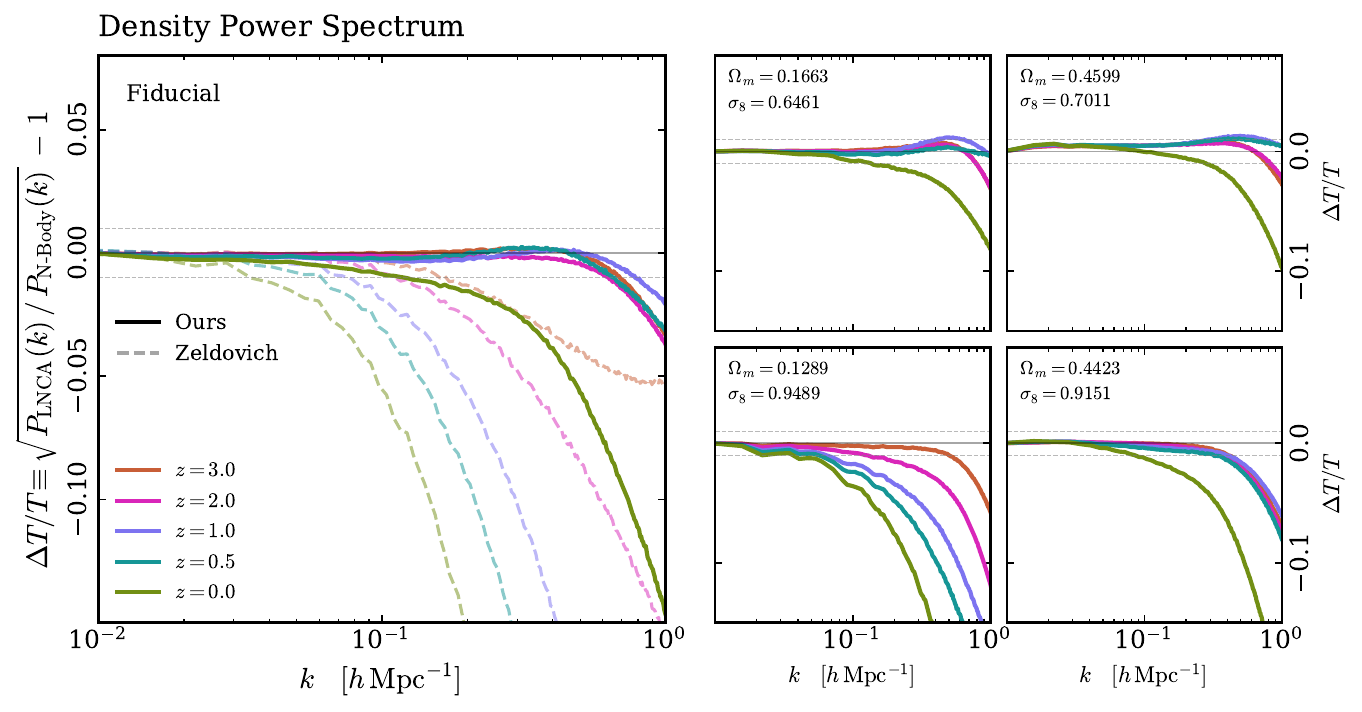}
  \caption{Power spectrum residuals represented as the relative transfer function $\Delta T/T$ for fiducial cosmology and four sample cosmologies from the Latin hypercube which were excluded from the training data. We see dramatic improvement in the spectral amplitudes at high $k$, with percent-level agreement for $k\lesssim 0.5 \, h \text{Mpc}^{-1}$.}
  \label{fig:pk}
\end{figure*}

\begin{figure*}
  \centering
  \includegraphics[width=0.99\linewidth]{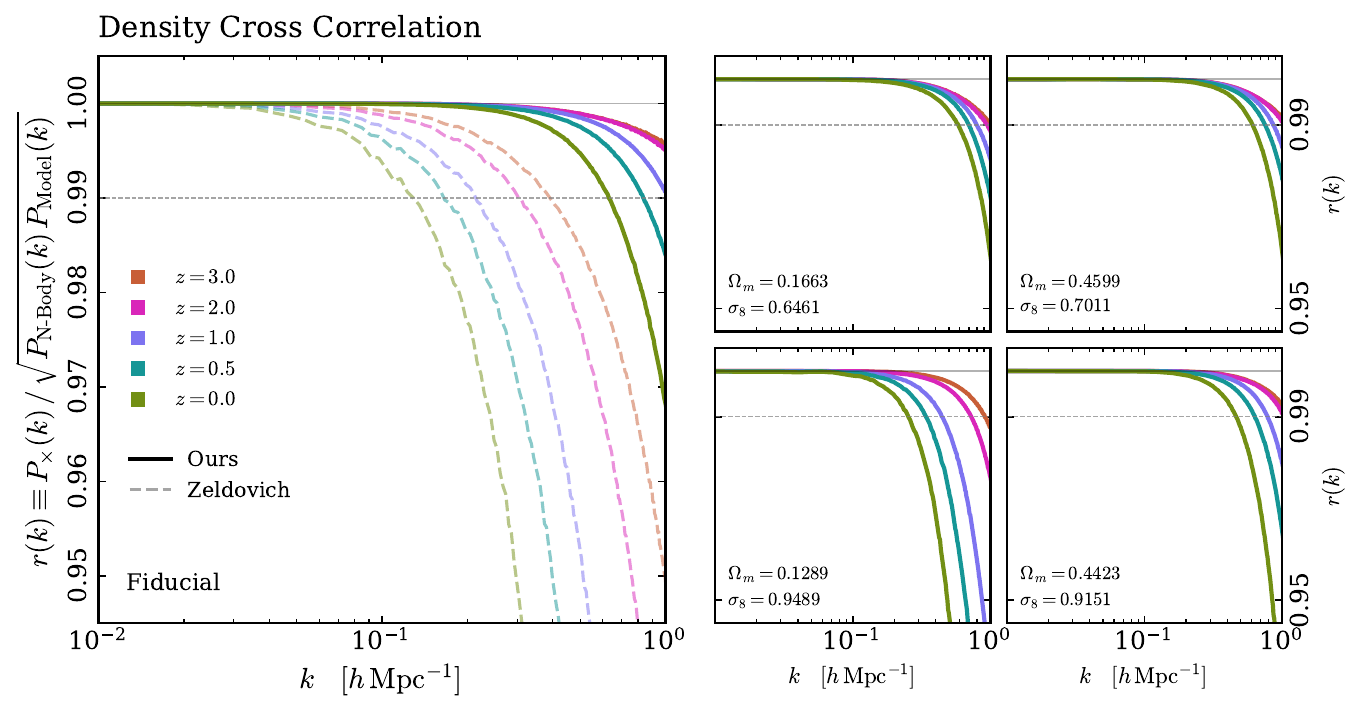}
  \caption{Cross-correlation residuals for fiducial cosmology and four sample cosmologies from the Latin hypercube which were excluded from training data. We see significant improvement in the phase alignment at high $k$, with percent-level agreement for $k\lesssim0.7 \, h \text{Mpc}^{-1}$.}
  \label{fig:r}
\end{figure*}

%\newpage

\subsection*{Power Spectrum}\label{sec:validation:power spectrum}

\noindent We compute the traditional matter overdensity field $\delta(\bm{x}) = \rho(\bm{x})/\bar{\rho} - 1$ from the evolved particle configurations using a Cloud-in-Cell (CIC) mass assignment onto a $512^3$ grid, and obtain the isotropic power spectrum $P_\delta(k)$ by spherically averaging the squared Fourier amplitudes of $\delta(\bm{x})$ in logarithmically spaced $k$-bins. To quantify the accuracy relative to the N-body reference, we define the fractional transfer deviation:

\begin{equation}
    \frac{\Delta T}{T}(k) \;\equiv\; \sqrt{\frac{P_{\mathrm{LNCA}}(k)}{P_{\text{N-body}}(k)}} - 1.
    \label{eq:DeltaT}
\end{equation}

Figure~\ref{fig:pk} shows the power spectrum residual for the fiducial cosmology across the full set of redshift snapshots used during training ($z \in \{3, 2, 1, 0.5, 0\}$). At $z=0$, our model achieves a power spectrum residual of $<1\%$ for all $k \lesssim 0.5\,h\,\mathrm{Mpc}^{-1}$, comfortably outperforming ZA across the same range by more than an order of magnitude. The accuracy degrades smoothly at higher $k$, with the model retaining sub-5\% deviations up to $k \sim 1\,h\,\mathrm{Mpc}^{-1}$. Higher redshifts exhibit notably better residuals at fixed $k$, consistent with the growth of mode-mode coupling under continued non-linear evolution. Figure~\ref{fig:pk} also extends to non-fiducial cosmologies drawn from the Latin Hypercube, spanning the \textit{extreme} (highly improbable) ranges of $\Omega_m$ and $\sigma_8$ explored during training and demonstrating the generality of our model.

\subsection*{Cross-Correlation}\label{sec:validation:crosscorr}

\noindent While $P_\delta(k)$ measures the recovered amplitude per mode, it is insensitive to mode-by-mode misalignment. To diagnose phase coherence directly, we define the cross-correlation coefficient:
\begin{equation}
    r(k) \;\equiv\; \frac{P_\times(k)}{\sqrt{P_{\mathrm{LNCA}}(k)\, P_{\text{N-body}}(k)}},
    \label{eq:rk}
\end{equation}
where $P_\times(k) \equiv \langle \delta_{\mathrm{LNCA}}(\bm{k})\,\delta^*_{\text{N-body}}(\bm{k})\rangle$ is the cross-power between the predicted and reference overdensity fields. By construction $r(k) \in [-1, 1]$, with $r(k) = 1$ corresponding to perfect mode-by-mode agreement and $1 - r(k)$ providing a clean measure of phase decoherence.

Figure~\ref{fig:r} shows $r(k)$ for the fiducial cosmology across the same five redshift snapshots. At $z=0$, our model maintains $1 - r(k) < 0.01$ for all $k \lesssim 0.7\,h\,\mathrm{Mpc}^{-1}$, demonstrating that the recovered power on these scales arises from correctly located Fourier modes rather than from amplitude-matched but phase-scrambled structure. 
%Relative to ZA, the learned correction preserves the relative phases of the underlying density field across more than a decade in $k$.
As with $P_\delta(k)$, performance degrades smoothly into the trans-linear regime, where $r(k)$ falls below the 1\% threshold but remains substantially above the ZA baseline. As expected, the model performs better for cosmologies closer to the fiducial center of its training distribution.

\subsection*{Halo Population}\label{sec:validation:halos}

\begin{figure}[b]
  \centering
  \includegraphics[width=\columnwidth]{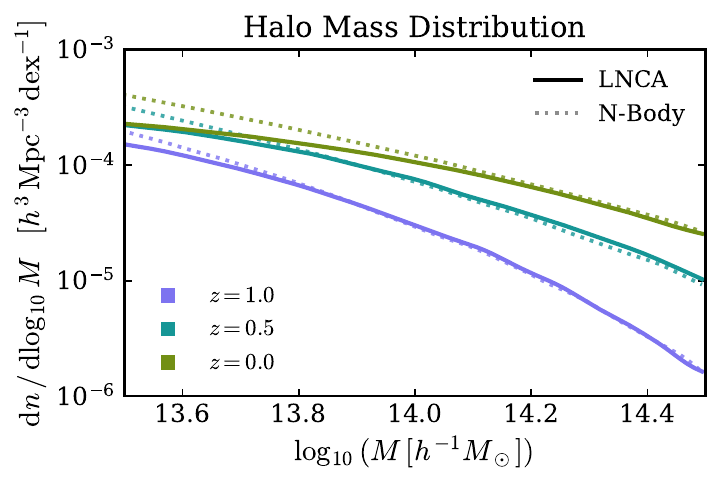}
  \caption{
    Halo Mass Functions generated by gaussian kernel density estimation for the target redshifts for our LNCA emulator vs the N-body ground truth at Fidicial Cosmology.
  }
  \label{fig:hmf}
\end{figure}

\begin{figure*}[thp]
  \centering
  \includegraphics[width=\textwidth]{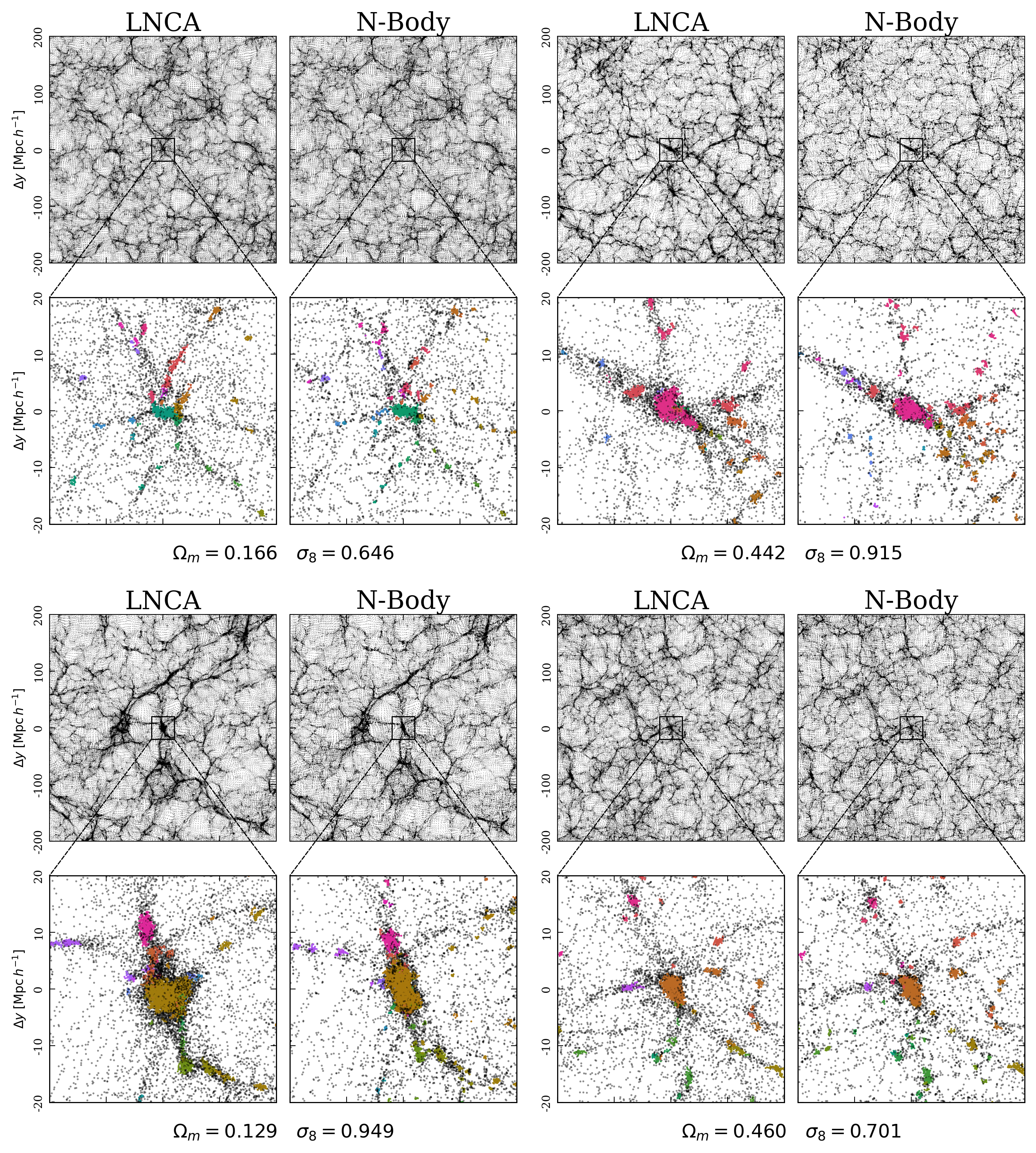}
  \caption{Visualization of the large halos collapsed by our model, compared to those present in the ground-truth N-body simulations. Each of the 4 quadrants corresponds to a cosmology within the Quijote Latin Hypercube which is not included in our training set and which is far from fiducial. Particles are colored if the FoF Halo-Finder identifies them as a member of a given halo, halo colors represent the bulk velocity direction of the particles in each halo. }
  \label{fig:LH_halo_comparison_4x4}
\end{figure*}

\noindent The diagnostics above constrain the field at the one- and two-point level, but neither verifies that the model collapses bound,
multi-particle objects of the correct abundance. Because the halo mass function
(HMF) is exponentially sensitive to the small-scale displacement field, it is
among the most stringent tests we can apply. We identify halos independently in
the predicted and reference fields with a friends-of-friends finder (linking
length $b_{\rm FoF}= 0.2$, minimum membership $N_{\rm min}=32$), and
compare the resulting mass functions in
Figure~\ref{fig:hmf}.

The agreement is excellent at the high-mass end: the abundance of the most
massive halos, built from the largest coherent infall regions, is recovered to
within the Poisson scatter of the reference at low redshift, confirming
that the explicit $\mathcal{L}_{\text{Halo}}$ term achieves its intended effect.
The agreement degrades toward the low-mass end, where the predicted abundance
falls short of the reference. This deficit shares a common origin with the
trans-linear power deficit: low-mass halos collapse from the
post-shell-crossing component of the displacement that carries near-zero
conditional expectation given the coarse inputs, so the network produces
slightly more diffuse objects that fall below the fixed FoF density threshold
more frequently than their N-body counterparts. 

The spatial fidelity of the recovered population is shown in
Figure~\ref{fig:LH_halo_comparison_4x4}, which compares the largest collapsed
halos in four held-out cosmologies from the corners of the Latin Hypercube, extremely far from the fiducial point.
Across all four the model places the massive halos at the correct locations and
reproduces their extent and size. The majority of satellite halos are reproduced as well, with a notable bias against the very smallest.
The residual disagreement is concentrated, as expected, in the cosmologies of highest non-linearity, where the abundance of collapsed structure is greatest and the per-particle displacement least predictable.

\begin{figure*}[t]
  \centering
  \includegraphics[width=0.99\linewidth]{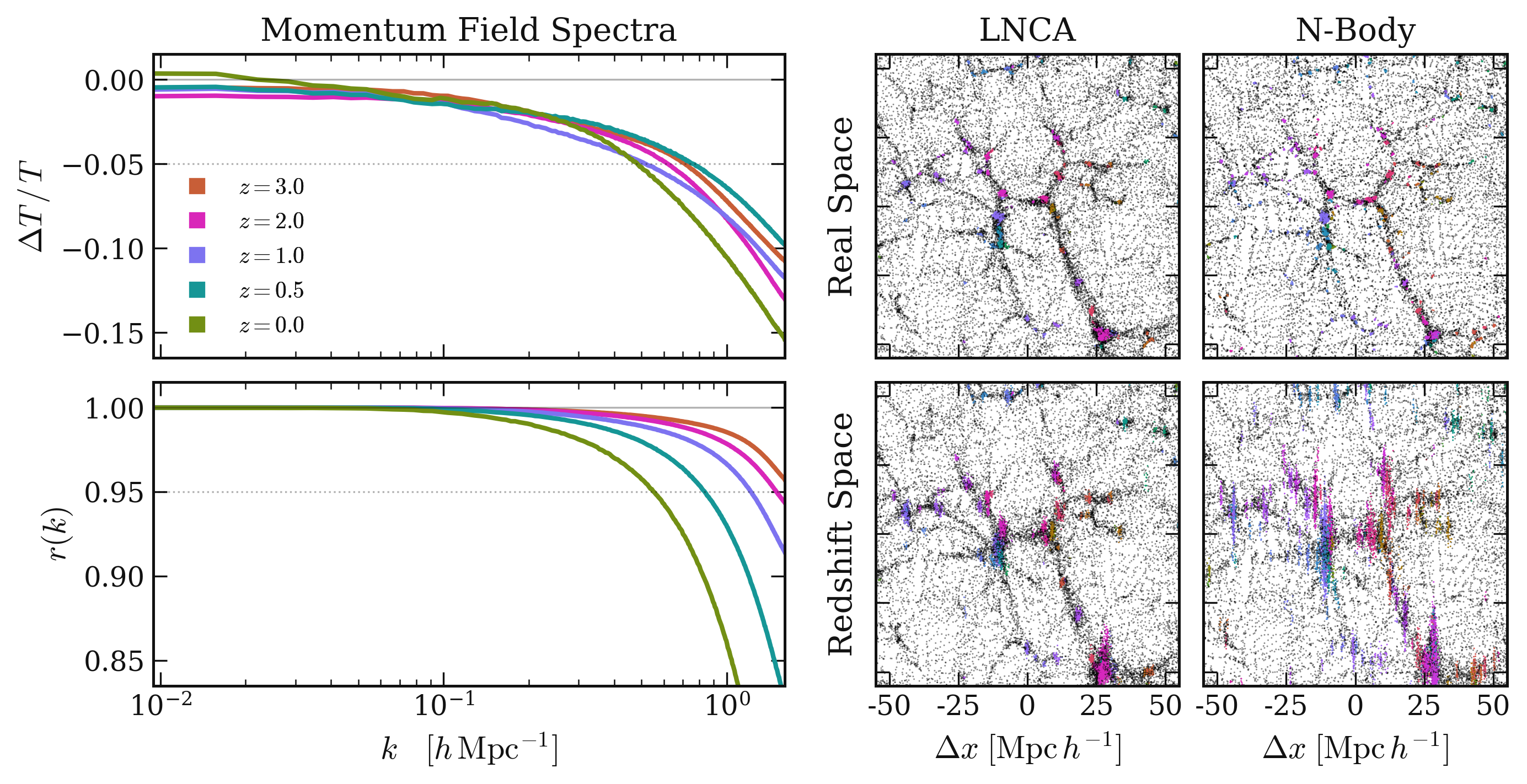}
  \vspace{-12pt}
  \caption{(Left) $\Delta T/T$ and $r(k)$ (eqs. \ref{eq:DeltaT}, \ref{eq:rk}) for the CIC-deposited momentum fields generated by our LNCA model relative to the target N-body. (Right) Redshift space distortions of our LNCA output particle field vs the target
    N-body at $z=0$.}
  \label{fig:Vel_RSD}
\end{figure*}

\newpage

\subsection*{Velocity Fields and \\ Redshift-Space Distortions}

\noindent
Finally, we display our model's performance in predicting the velocity fields and their downstream observables. The individual particle velocities are output directly by our model and can be extracted as the time derivative of the particle displacements, $\dot{\bm{\Psi}}_{\mathrm{LNCA}}$, at each timestep, before they are passed to the integrator to get $\bm{\Psi}$. Even with our momentum-specific loss term, the individual particle velocities are less constrained by the training process due to their intermediate nature in the integration; many particle trajectories can lead to the same final particle positions, let alone the same density fields. Given this, and the highly unpredictable and extreme-valued velocities found within halos, it is expected that the accuracy on velocity reconstruction will be substantially poorer. 

To quantify the impact on our observables, we show the residuals of the output momentum density field in Fourier Space, relative to the Target N-body. The momentum density field is constructed by CIC-depositing each particle's velocity onto the Eulerian mesh at each redshift snapshot, weighted by particle mass. This yields a vector field which can be decomposed into three scalar field components. The spectral quantities are calculated over these scalar component fields and averaged to get the momentum-space transfer function $\Delta T/T(k)$ and cross-correlation $r(k)$ shown above in the left panel of Fig. \ref{fig:Vel_RSD}. $\Delta T/T(k)$ and $r(k)$ remain within $5\%$ of the target for $k \lesssim 0.5\,h\,\mathrm{Mpc}^{-1}$ at all redshifts, before degrading sharply to $25\%$ by $k \simeq 1\,h\,\mathrm{Mpc}^{-1}$.

\newpage

The right panel of Fig.~\ref{fig:Vel_RSD} shows the corresponding particle fields in redshift space.
On large scales our model accurately reproduces the coherent infall which compresses structures
along the line of sight. However, as expected, the small-scale distortions are not recovered. In the N-body reference, virialized halos appear as ``fingers of God'', elongated along the line of sight by internal velocity dispersions that support halo morphology and are uncorrelated with the bulk flow. Our emulator does not capture these virial velocity dispersions and thus systematically underestimates this elongation, leaving collapsed objects closer to their real-space morphology than they should be. 

The origin of this deficit is due to the same post-shell-crossing internal phase-space discontinuity that suppresses the formation of halos in the first place. The particle-wise sub-halo dynamics lie beyond both the Lagrangian receptive field of our LNCA and the conditional expectation a deterministic $L_2$ objective can represent. The explicit $\sigma_{p,h}$ moment in $\mathcal{L}_{\rm Halo}$ (Eq.~\ref{eq:lhalo}) targets this dispersion, but the stability of the other training objectives actively selects against the extreme-valued velocity dispersion within halos. Thus, the predicted velocities do not reproduce the full profile that sets the finger length. The LNCA
velocity field is therefore suitable for observables weighted toward coherent flows, including linear RSD multipoles, pairwise velocities, and kSZ, while small-scale redshift-space clustering would require the dispersive component to be supplied separately.

\newpage

\section{Discussion}

\noindent The validation metrics presented above establish that our Lagrangian Neural Cellular Automaton (LNCA) reproduces the non-linear matter field with percent-level fidelity across a broad range of scales, redshifts, and cosmologies. We now situate these results within the landscape of existing emulation methods, both to calibrate our model's accuracy against the current state of the art and to argue that its principal contributions lie not in raw accuracy but in its physically constrained structure, efficiency, and interpretability.

\subsection*{Comparison to Existing Emulators}

\noindent The most direct point of comparison is the field-level emulator of Jamieson et al.~\cite{Jamieson2025}, which, like our model, learns a residual correction to the Zeldovich displacement field on the Quijote Latin Hypercube and conditioned on cosmology and time. Their U-Net achieves percent-level accuracy in both amplitude and phase up to $k \sim 1\,h\,\mathrm{Mpc}^{-1}$ at $z=0$, whereas our model maintains $|\Delta T/T| < 0.01$ and $1 - r(k) < 0.01$ to $k \sim 0.5\,h\,\mathrm{Mpc}^{-1}$, with performance degrading smoothly beyond this scale. The two models are comparable in their recovery of halo populations (Fig.~\ref{fig:LH_halo_comparison_4x4}) and in their component-wise displacement residuals (Fig.~\ref{fig:delta_disp}). This near-parity is maintained not only at the fiducial cosmology but across the extremes of the Latin Hypercube, indicating that both models interpolate the cosmology dependence well across most of the hypercube, with degradation confined to its extreme corners. Our emulator therefore approaches, though does not yet match, the accuracy reach of the leading convolutional baseline, while substantially exceeding the fidelity of the COLA method~\cite{Tassev2013} from which our residual formulation is inspired, and of earlier Eulerian CNN approaches~\cite{Ramanah2019} that lack a matter-conserving Lagrangian advection mechanism.

This near-parity in accuracy is achieved with radically reduced model complexity. The densely connected U-Net of \citet{Jamieson2025} requires on the order of $3\times10^{6}$ learnable weights, distributed across a hierarchy of convolutional channels and resolution scales, to attain its reported accuracy. By contrast, our recurrent construction learns an \textit{emergent} dynamical rule parametrized by only $\sim 4\times10^{3}$ terms, a reduction of nearly four orders of magnitude. The accuracy we obtain is not encoded statically in a large bank of filters applied once, but emerges from the repeated application of a single, compact update rule over the integration rollout. This recurrence, in conjunction with the stringent constraints we place on continuity and equivariance, naturally yields this kind of dramatic improvement in parameter economy. We consider this demonstration that a significantly reduced model size can support emulation of comparable fidelity to be a central finding of this work.

\newpage

The two fast solvers most relevant to our regime, \textsc{FastPM}~\cite{Feng2016} and
\textsc{pmwd}~\cite{Li2022}, occupy complementary positions that the present work does not
displace. \textsc{FastPM} is a CPU code whose two-dimensional domain decomposition scales to
large processor counts, designed for mass production of approximate realizations where
throughput at fixed modest accuracy determines utility. \textsc{pmwd} is a single-GPU
differentiable library whose adjoint formulation eliminates the
$\mathcal{O}(N_\mathrm{steps})$ memory cost of reverse-mode differentiation, which suits
gradient-based field-level inference but bounds the accessible volume by the memory of one
accelerator, with $512^3$ particles already near that bound. Both are as fast as our emulator
at fixed volume, requiring $\sim 10-20$ seconds to evolve $512^3$ particles to $z=0$ on a single A100 (40 GB) GPU. 
We therefore claim no advantage in wall-clock time. 
We claim instead that the properties these codes trade against one another, parallel
scalability in one case and differentiability with small-scale fidelity in the other, may both be achieved when the update rule is local.

That trade-off originates in the Fourier Transform. Solving Poisson's equation spectrally is
irreducibly global, since the density must be transformed along all three axes and no rank
holds a complete row of the mesh, so data must be redistributed between successive sets of
one-dimensional transforms. The communication volume is the
entire mesh twice per step, once for the density and once for the force, and it imposes a
synchronization barrier that cannot be overlapped with computation, because the potential at
any point depends on the density everywhere. Beyond a certain volume this global traffic restricts the wall-clock time more than the arithmetic itself. 
A local update rule admits no such barrier. The LNCA propagates information at finite speed $c$ per step, so each
subvolume requires only a halo of depth $c$ from its neighbors, and communication scales with
surface area rather than volume, is nearest-neighbor rather than global, and overlaps with the
interior update.

Locality also naturally yields a capability that one-shot emulators lack. Because the model integrates a
trajectory rather than evaluating a single map, positions and velocities exist at every step of
the rollout, and a particle can be frozen when it crosses the observer's past lightcone. 
A network mapping linear to nonlinear displacements directly returns a snapshot at the redshift set by its style parameters, so lightcones remain accessible but their cost grows with the number of radial shells, each
requiring an independent pass over the full volume. 
Our rollout visits every intermediate time already, so lightcone output incurs no additional passes, and finite $c$ ensures its causal structure extends onto to a small local subvolume. 
The case for the LNCA is therefore not that it is the
fastest forward model available, but that locality and recurrence supply scalability,
differentiability, and lightcone construction within a single architecture.

\newpage

\subsection*{Physical Structure and Interpretability}

\noindent We argue that the central advantage of our framework is not its accuracy relative to the convolutional state of the art, but the physical constraints, interpretability, and explicit internal dynamics it provides. Existing field-level emulators have typically relied on dense convolutional networks with wide receptive fields and opaque internal representations. Such architectures impose few of the physical symmetries that govern gravitational structure formation: they are not typically rotationally equivariant (with some exceptions, see \cite{Dai2022, thiele2022}), and their stacked downsampling and upsampling operations extend the receptive field to a large fraction of the box, so the support of the learned map is fixed by architecture, with no principled bound on the rate of causal influence.

Our model is designed intentionally to be strictly local, equivariant, and smoothly continuous in time. By expressing every learned update as a linear combination of an equivariant vector menu modulated by invariant scalar coefficients, the LNCA is $E(3)$-equivariant by construction rather than by data augmentation. 
The cellular-automaton substrate restricts information transport to a single cell per substep, so the region of the Lagrangian lattice on which any output depends is finite, known, and grows linearly with the rollout. We stress that this is a property of the emulator and not a symmetry of the system it emulates. 
%The target dynamics are Newtonian gravity on an FLRW background, in which the potential is recovered from an instantaneous Poisson solve rather than a retarded propagator; every region influences every other without delay, and relativistic corrections enter only at near-horizon scales absent from the simulations used here. The physically relevant quantity is instead the \emph{support scale} of the nonlinear map: the radius in q within which the initial field determines the late-time nonlinear displacement of a Lagrangian patch. This scale is finite and modest, set by the distance material travels before joining the structure it ends up in, which is why a bounded receptive field suffices for convolutional emulators as well as for ours.
Furthermore, because the network predicts the time derivatives of the displacement and latent fields, which are then integrated by a symplectic scheme, our model generates trajectories that are smooth and continuous in time by construction, as the solution to a system of coupled differential equations. 

\begin{figure*}[thp]
  \centering
  \includegraphics[width=\textwidth]{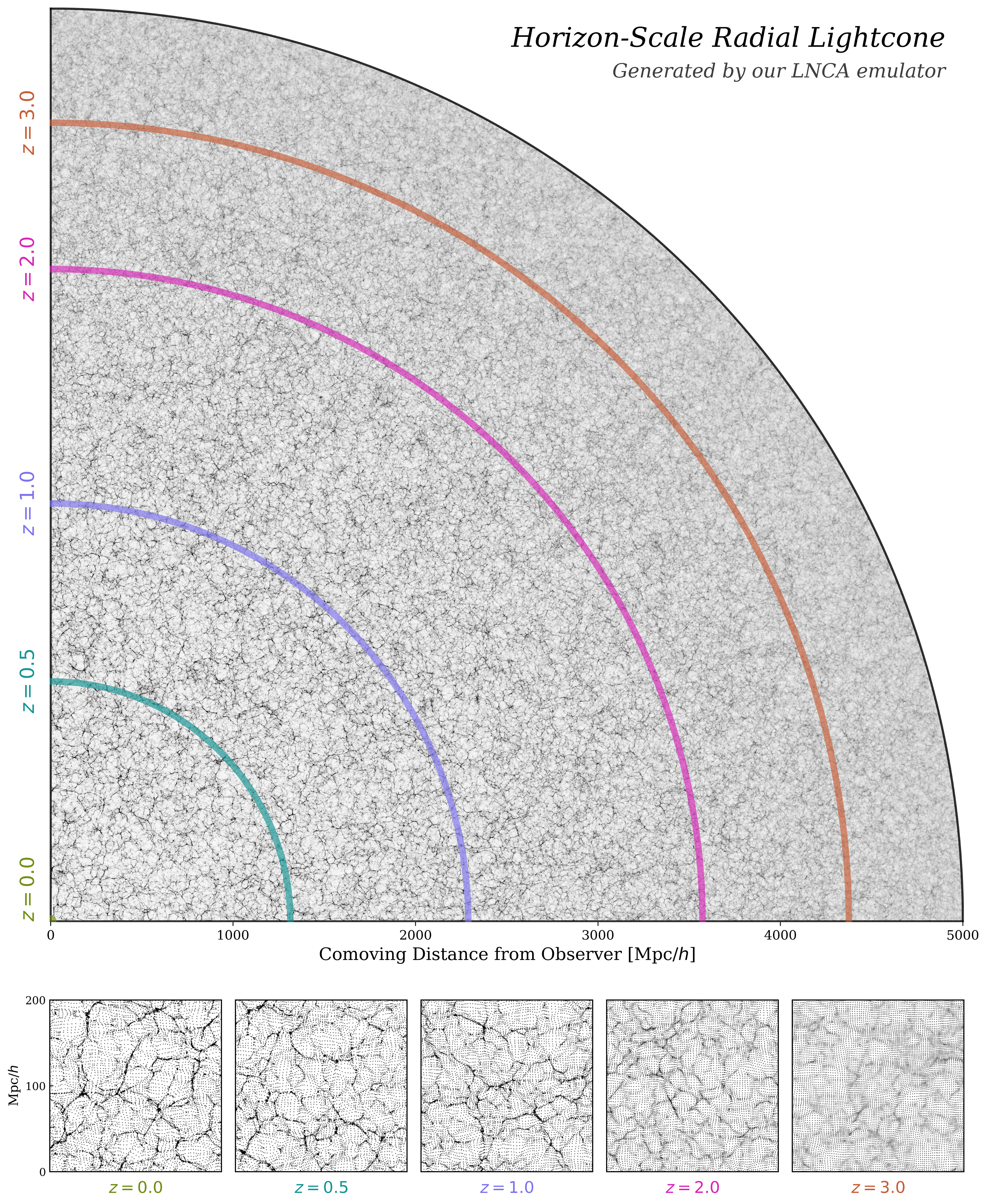}
  \caption{Example ``lightcone'' generated by our LNCA forward model. This volume features an observer at the origin and demonstrates the ability of our model to naturally respect the geometry of the temporal horizon. The apparent redshift of each particle is a continuous function of its distance to the origin, yielding a volume in which structure appears less collapsed further from the observer. The lower row highlights $200 \,\text{Mpc}\,h^{-1}$ wide crops of the above volume to demonstrate the effect.
  }  
  \label{fig:lightcone}
\end{figure*}

This last property carries practical consequences for survey modeling. A U-Net emulator that maps initial conditions directly to a fixed output redshift must be re-evaluated independently for each desired epoch and interpolated to yield realistic survey geometry. Our integrator instead produces a single continuous trajectory, from which any intermediate state may be read off. This is precisely the structure required to construct \textit{lightcones}: mock realizations which explictly encode the horizon geometry, reproducing the observational feature that structures at greater comoving distance are seen at earlier cosmic times. The continuous, point-wise-timestamped evolution of the LNCA supports this lightcone construction natively (as shown in Fig.~\ref{fig:lightcone}), whereas the snapshot-based convolutional approach does not.

Finally, our architecture yields to direct inspection of its internal dynamics. Because the state of each node comprises explicit latent scalar and vector fields evolving under an analytic, polynomial update rule, we may examine the behavior of these latent channels over the course of the rollout and relate them to the physical deformation history they encode. We present such an analysis in Appendix \ref{app:interp}, where the emergence of structure in the latent fields offers a window into the learned dynamics that is absent from any existing models.

\newpage

\subsection*{Limitations of our Model}

Our model shares a variety of shortcomings with existing field-level emulators such as \citet{Jamieson2025}. The model is exclusively trained on $1\,\mathrm{Gpc}/h$ configurations of $512^3$ particles, so transferability to other resolutions, with their attendant change in effective force resolution, remains untested. It is also a purely gravitational emulator and omits baryonic and sub-grid physics, which limits direct comparison to observations. These are common to the current generation of learned N-body surrogates rather than specific to our approach.

An additional class of limitation is more intrinsic to our formulation, arising as the direct cost of the very physical constraints we describe above. Most consequential among these is the impact of bounded information propagation on virialized structure. Because the LNCA communicates over a fixed neighborhood in Lagrangian $q$-space, information can traverse only a finite distance per substep, and the total receptive field accumulated over the integration is correspondingly bounded. This effect is most noticeable inside collapsed halos which bring together material from disjoint regions of the initial density field. The finite Lagrangian receptive field therefore imposes a hard ceiling on the recoverable internal halo morphology as the relevant non-local information is simply not accessible within the neighborhood over which the automaton communicates.

The extent of this horizon is in turn constrained by our requirement of continuous dynamics: because we train through the full integration from $z=127$ to $z=0$, backpropagation through time (BPTT) must retain the intermediate state across all substeps, and the associated memory cost bounds the substep count and neighborhood size during training. The information horizon and the morphology ceiling it induces are thus not independent failures but downstream consequences of the same design choices that make the model physically constrained and its dynamics smooth.

We emphasize that our construction is a substantial departure from established practice. Convolutional U-Net emulators inherit years of accumulated architectural refinement, tooling, and training heuristics from the broader machine-learning literature, whereas the LNCA is novel in nearly every component, and we accordingly do not expect it to match such highly optimized baselines at this early stage. We regard the present results as a demonstration that a compact, equivariant, causally local model can approach that accuracy at all, and we expect that several of these limitations are amenable to improvement with further experimentation, whether through enlarged or multi-scale neighborhoods, hierarchical message passing that extends the effective horizon at fixed local cost, memory-efficient adjoint schemes that relax the BPTT constraint, or simply the more mature training methods that such architectures will accrue over time. We regard their systematic exploration as a promising direction for future work.

\newpage
\section{Conclusion}\label{sec:conclusion}

\noindent In this work, we have presented the LNCA, a novel forward model and field-level emulator for cosmic structure formation with a number of desirable properties not present in existing models. Our model is a type of localized, recurrent graph neural network which operates in the Lagrangian frame. We train this model to transform Zeldovich particle displacements into corresponding non-linear displacement residuals, given a target redshift and cosmology, using a suite of N-body simulations. We then validate our trained model on a variety of metrics, relative to held-out N-body simulations with both fiducial and extreme cosmologies. 

Our model leverages its recurrent structure to generate particle trajectories forward in time over 32 discrete timesteps from $z=127 \rightarrow z=0$ by predicting each particle's acceleration as a function of its local neighborhood and iteratively producing the next field state via a symplectic integrator. The internal learned function trained to predict this is a purely analytic Kolmogorov-Arnold  Network (KAN) consisting entirely of learnable Taylor expansions. The combination of these properties ensures that the output dynamics is smooth and continuous in time. The network is further constrained to operate entirely in terms of equivariant features of a set of latent fields, which guarantees rotationally and translationally equivariant behavior.  

Our emulator achieves percent-level accuracy in the amplitudes and phases of the matter density field at $k \lesssim 0.5$ for $z=0$, as shown in Figs.~\ref{fig:pk} \& \ref{fig:r}, with better performance at higher redshifts where the non-linear dynamics are less pronounced. These validation results are comparable to existing emulators \cite{Jamieson2025}, while offering continuous trajectories generated from simple and interpretable latent field dynamics.

%\newpage

%Beyond these quantitative benchmarks, we wish to emphasize that the central contribution of this work is methodological. While existing field-level emulators have achieved their accuracy through dense convolutional networks with millions of opaque parameters fields, our model attains comparable fidelity with a compact, analytic update rule of only a few thousand terms, while imposing $E(3)$ equivariance, restricting information propagation, and generating continuous trajectories. 
%These properties distinguish the LNCA as a transparent and physically constrained dynamical model whose internal behavior can be trusted and interrogated. 

Beyond these quantitative benchmarks, we wish to emphasize that the central contribution of this work is methodological. While existing field-level emulators have achieved their accuracy through dense convolutional networks with millions of opaque parameters, our model attains comparable fidelity with a compact, analytic update rule of only a few thousand terms, while imposing $E(3)$ equivariance, restricting information propagation, and generating continuous trajectories. As the next generation of surveys drives cosmology toward full field-level reconstruction and simulation-based inference, where the forward model must be evaluated many thousands of times within a differentiable optimization or sampling loop, we expect that models of this kind, which are cheap, continuous in time, local by construction, and faithful to the symmetries of the underlying physics, will prove increasingly valuable. In the remainder of this section we outline the applications these properties are intended to enable.

\subsubsection*{Future Work}

\noindent The most natural and immediate application of the LNCA is the mass production of survey-quality mock catalogues. Because the model is cheap in both time and memory and produces continuous, point-wise-timestamped trajectories, it is well suited to generating the large ensembles of lightcone realizations required to characterize covariance matrices and observational systematics for stage-IV surveys, at a tiny fraction of the cost of full $N$-body simulations. The continuous evolution is particularly advantageous for probes that integrate along the line of sight rather than sampling a single epoch. For weak-lensing surveys such as LSST and \textit{Euclid}, the native lightcone construction supports the direct generation of convergence and shear maps, capturing the full redshift dependence of the lensing kernel without interpolation between discrete snapshots.
Our model is also well-suited for generating mocks of the kinetic Sunyaev-Zel'dovich (kSZ) effect, whose signal depends on the product of the electron density and the line-of-sight peculiar velocity. Because our model evolves the velocity field jointly and self-consistently with the density, we naturally enable the construction of mock kSZ observables. 

More ambitiously, the differentiability and efficiency of our model enables its application to field-level inference and reconstruction. The LNCA is designed to be suitable as the forward operator within a gradient-based sampler to reconstruct the initial conditions and cosmological parameters jointly from observed galaxy, lensing, and/or kSZ fields. Its compact, differentiable construction is precisely what makes the many-thousand-evaluation inner loops of such inference tractable, and we regard the coupling of the LNCA to a full field-level inference pipeline, validated against these upcoming datasets, as the principal direction of our ongoing work.

\section*{Acknowledgments}
{\scriptsize \noindent The authors are grateful to Tom Abel, Marcelo Alvarez, Adrian Bayer,  Josh Frieman, Ben Horowitz, Mikhail Ivanov, Drew Jamieson, Elisabeth Krause, Jia Liu, Zarija Lukic, Noah Sailer, Emmanuel Schaan, Uros Seljak, Marko Simonovic, Leander Thiele, Francisco Villaescusa-Navarro, Ben Wandelt, Risa Wechsler and others for their constructive feedback and insightful suggestions.

CJ ~gratefully acknowledges fellowship support from the National Science Foundation's Graduate Research Fellowship Program (GRFP) and Stanford University's William R. Hewlett Graduate Fellowship. OHEP is supported in part by the National Science Foundation under grant number PHY-2609396.

This work utilized computational resources from the National Energy Research Scientific Computing Center (NERSC), a user facility supported by the Dept. of Energy Office of Science.  

Meaningful software development was performed using Large Language Models, Claude Sonnet 4.6 and Opus 4.8, which were made available to the authors through Stanford University's enterprise subscription. All LLM-generated code was verified and extensively tested by the authors. \textit{No LLM-based tools were used to interpret results or write text for this document.}

\textcolor{white}{Vertical whitespace}
}

\newpage

\newpage

\section*{Appendix}

\subsection{Further Training Details}\label{app:training}
\noindent 
The model structure we propose is unusual in its radical simplicity and the emergent nature of its learned mechanics. Unlike traditional deep convolutional networks, which distribute learned behavior across millions of disparate parameters, we require accurate dynamics to emerge from a small family of interacting differential equations. The recurrent nature of the model generates an exceptionally rough, fractal loss landscape which gradient methods alone struggle to navigate. We therefore employ a coupled evolutionary scheme, in addition to gradient descent, allowing a set of models to compete, mutate, and expand while balancing performance against model complexity. 

\subsubsection*{Sparsity Optimization}
\noindent A central ambition of our project is to produce a model that is as simple as possible in the sense of the volume of learned parameters. The recurrent, symmetric, point-wise definition already massively reduces complexity; to further discover the appropriate network dimensions we add a sparsity term to the physical loss terms of \S\ref{sec:model_design} that penalizes spurious parameter usage and encourages the model to self-prune useless pathways as training progresses. We use a Group-Lasso regularizer for this purpose. Where the ordinary Lasso adds an $L_1$ penalty on individual weights and drives scattered elements to zero, the Group-Lasso first partitions the weights into predefined groups $g$ and penalizes the sum of their Euclidean norms, $\lambda\sum_g \sqrt{p_g}\,\lVert \mathbf{w}_g\rVert_2$, where $\mathbf{w}_g$ is the weight vector of group $g$, $p_g$ its size, and $\sqrt{p_g}$ a factor that equalizes pressure across groups of unequal dimension. Because the $L_2$ norm of a group is non-differentiable only at the origin, minimizing it pushes an entire group to zero simultaneously rather than thinning it, yielding sparsity at the level of whole structures instead of stray weights. We define one group as all weights downstream of each input feature, so that an unused invariant is driven to zero coherently and can be removed from the graph outright. Once a group norm falls below a fixed threshold, $\tau_{\rm prune}=0.01$, the pathway is hard-pruned and frozen at zero, converting soft regularization into a discrete reduction in size and cost. The penalty is held inactive for the first 200 epochs of each generation so promising pathways can establish a gradient signature before becoming eligible for removal. The sparsity pressure is deliberately weak ($\sim 1\%$ loss pressure), so feature selection proceeds gradually, and the surviving invariants give a direct, interpretable readout of which initial-condition quantities the dynamics actually require.

\newpage
\subsubsection*{Evolutionary Mechanics}
\noindent 
Rather than expecting immediate convergence to a robust rule set, we let the model slowly evolve and expand its complexity, maintaining a population of 16 candidates and alternating an inner loop of gradient refinement with an outer loop of selection, mutation, and expansion.

\vspace{12pt}
\noindent 
\textbf{Inner-loop training:} Within each generation, every individual is refined for $1000$ epochs of gradient descent, supplying the local exploitation that settles a model into the nearest basin of the fractal landscape while the outer loop handles global exploration. We use the popular \texttt{Adam} optimizer with a cosine learning rate schedule, chosen for its stability across the poorly-conditioned curvature that recurrence produces and for requiring no per-generation retuning. Newly added parameters (see \emph{Model expansion}) enter with fresh optimizer state and adapt over the remaining epochs of the generation in which they are born.

\vspace{12pt}
\noindent 
\textbf{Multi-objective selection:} Between generations, 4 survivors are chosen by non-dominated sorting with crowding-distance tie-breaking (NSGA-II), which keeps the retained front spread across the objective simplex rather than collapsing onto a single scalarization. The Pareto front is defined over the displacement, momentum, and real-space density objectives $(\mathcal{L}_\Psi,\ \mathcal{L}_{\mathbf p},\ \mathcal{L}_\delta)$. The surviving models are thus a meaningfully diverse set of the highest performers. These survivors are then the seeds for the next generation of models, which are initialized from random offsets of the parents. %At generation zero, however, we rank on $\mathcal{L}_\Psi$ alone: while models are still non-viable these objectives are strongly correlated ($\rho > 0.9$), so the momentum and density terms carry little independent ranking information and are gated into the selection criterion only from generation one, once individuals are good enough to express genuine trade-offs among them.

\vspace{12pt}
\noindent 
\textbf{Model expansion:} Complexity grows for new generations through a mutation we term \emph{``field birth''}, which introduces a new latent field and its interaction operator into a selected individual. The output coupling is zero-initialized so the new field contributes exactly nothing to every objective at the moment of insertion, while the input coupling is left live so that gradients flow into the pathway immediately and it can begin to specialize. This makes expansion strictly loss-preserving, which is essential under NSGA-II, where a transient penalty for growing would cull an otherwise promising lineage before its added capacity could pay off. Growth of this kind is well suited to emergent problem solving precisely because the correct model size is not known a priori: rather than fixing an architecture and fitting it, we add capacity only where a live gradient signal rewards it, while the Group-Lasso pressure of the previous section removes whatever capacity is not contributing. The two act as opposing forces over the run, letting the population search the joint space of architecture and dynamics; the final model is the best performing on the combined metric $\mathcal{L}_\Psi+ \mathcal{L}_{\mathbf p}+\mathcal{L}_\delta + \mathcal{L}_{\rm Halo}$ after 32 generations, yielding the compact ($\sim\!4{,}000$-parameter) network described in the main text.

\newpage

\subsection{Trained Model Structure}

Because the update rule $f_\theta$ is a compact KAN acting on a pool of
rotationally invariant scalars and emitting scalar weights on a fixed menu of
equivariant vectors, the trained model admits a direct read-out of what it has
learned to use. We quantify this with two attributions, pooled across the full rollout and
weighted by the per-step correction $\langle|\ddot{\Psi}_{\rm res}|\rangle$.
For each input invariant we report a history-weighted whole-network sensitivity
$\Sigma_j$ (how strongly the output depends on that feature) and a coupling
fraction $\Upsilon_j\in[0,1]$ (the share of that dependence carried by
interaction with other features rather than by a standalone response). For each
menu vector we report a signed contribution $c_k$ (its projected share of the
net acceleration, summing to unity) and a coherence
$\bar{\alpha}_k\in[-1,1]$ (its mean alignment with the net update, so that a
large-magnitude vector held in near-cancellation registers as low $\bar{\alpha}_k$).
The scalar-pool invariants are strongly correlated so scores may distribute credit among correlated features.

The two strongest invariants are the displacement shear amplitude
$|\mathbf{S}_\Psi|^2$ and the tidal projection of a learned latent field gradient
$\nabla\mathcal{S}_1\cdot(\mathbf{T}_\Psi\cdot\nabla\mathcal{S}_1)$, suggesting the learned corrections are governed
mainly by the local anisotropy of collapse and by how the accumulated scalar
memory aligns with that anisotropy, and both operate through mode coupling
rather than as standalone terms. The typically-high value of the coupling term $\Upsilon$ indicates that the model's behavior is generated from a high degree of interaction, rather than independent feature injection. 
Growth enters through $D(t)$ and 
its high coupling ($\Upsilon_j=0.77$) indicates that $D(t)$ is used to modulate
the strength of the structural corrections.

The synthesized acceleration is built from a few coherent directions. The
residual velocity $\dot{\Psi}_{\rm res}$ and the two latent vector
channels $\mathcal{V}_0,\mathcal{V}_1$ supply most of the update. 
The residual Laplacian $\nabla^2\Psi_{\rm res}$ contributes
appreciably but at low coherence to cancel/smooth infall. The
exact 2LPT velocity template $\dot{\Psi}^{(2)}$ is nearly unused, even though the 2LPT acceleration magnitude ranks
highly as an input feature.

Both tables use the following notation.
$\mathcal{S}_i$ and $\mathcal{V}_i$ are the $i$th latent scalar and vector
channels, with traceless (tidal) Hessian $\mathbf{H}[\mathcal{S}_i]$ and
vorticity $\boldsymbol{\omega}_i \equiv \nabla\times\mathcal{V}_i$. A vector
gradient splits into a symmetric strain and an antisymmetric vorticity part;
for the displacement $\Psi$ we keep the strain $\mathbf{S}_\Psi$ and its
traceless tidal part $\mathbf{T}_\Psi$ (both symmetric), and for the residual
velocity $\pi\equiv\dot{\Psi}_{\rm res}$ the strain $\mathbf{S}_\pi$ and
vorticity $\boldsymbol{\omega}_\pi$.
The deformation Jacobian $\mathbf{J}_\Psi \equiv \partial\mathbf{x}/\partial\mathbf{q} =
\mathbf{I} + \nabla_q\Psi$ has a vanishing determinant at shell crossing, hence the $\tanh$ compression.
$\Psi^{(1)}$ and $\dot{\Psi}^{(1)}$ are the linear displacement and
velocity while $\Psi^{(2)}$, $\dot{\Psi}^{(2)}$,
$\ddot{\Psi}^{(2)}$ the exact second-order LPT templates and
$S^{(2)}_{\rm src}$ the 2LPT source scalar. $a$ is the scale factor, $D(a)$
the linear growth factor, and $f(a)=\mathrm{d}\ln D/\mathrm{d}\ln a$ the growth rate. Double dots denote contraction, $A\!:\!B \equiv \sum_{ab}A_{ab}B_{ab}$.

\newpage

\begin{table}
\caption{Relevance of the input scalars, averaged over the full rollout ($a=1/128\to1$) and weighted by per-step corrective work.
$\Sigma_j$: history-weighted whole-network gradient sensitivity, normalized to
its maximum. $\Upsilon_j\in[0,1]$: coupling fraction, the share of feature $j$'s
effect variance carried by interaction with other features rather than its
standalone main effect. }
\label{tab:invariant_relevance}
\begin{ruledtabular}
\begin{tabular}{lcc}
Input Features & Sensitivity $\Sigma_j$ & Coupling $\Upsilon_j$ \\
\hline
$\nabla\mathcal{S}_1\cdot(\mathbf{T}_\Psi\cdot\nabla\mathcal{S}_1)$ & $1$ & $0.73$ \\
$|\mathbf{S}_\Psi|^2$ & $0.95$ & $0.73$ \\
$\nabla^2\mathcal{S}_1$ & $0.86$ & $0.69$ \\
$|\ddot{\Psi}^{(2)}|^2$ & $0.82$ & $0.5$ \\
$\mathcal{S}_1$ & $0.75$ & $0.18$ \\
$S^{(2)}_{\rm src}$ & $0.74$ & $0.7$ \\
$(\dot{\Psi}_{\rm res})\cdot(\nabla^2\Psi_{\rm res})$ & $0.72$ & $0.95$ \\
$\mathbf{T}_\Psi\!:\!\mathbf{H}[\mathcal{S}_1]$ & $0.71$ & $0.12$ \\
$(\dot{\Psi}_{\rm res})\cdot(\mathbf{T}_\Psi\cdot\nabla^2\Psi_{\rm res})$ & $0.67$ & $0.95$ \\
$|\dot{\Psi}^{(1)}|^2$ & $0.63$ & $0.72$ \\
$\tanh(\det\mathbf{J}_\Psi)$ & $0.57$ & $0.85$ \\
$\nabla^2\mathcal{S}_0$ & $0.56$ & $0.86$ \\
$\det\mathbf{S}_\Psi$ & $0.56$ & $0.82$ \\
$|\Psi^{(2)}|^2$ & $0.56$ & $0.41$ \\
$\nabla\cdot\dot{\Psi}_{\rm res}$ & $0.53$ & $0.84$ \\
$D(t)$ & $0.45$ & $0.77$ \\
$f(a)$ & $0.28$ & $0.71$ \\
$a$ & $0.14$ & $0.49$ \\
\end{tabular}
\end{ruledtabular}
\end{table}

\begin{table}
\caption{Menu contribution to $\ddot{\Psi}_{\rm res}$, pooled as in
Table~\ref{tab:invariant_relevance}. $c_k$: history-integrated signed projection
share, summing to 1 across the menu. $\bar{\alpha}_k\in[-1,1]$: alignment, the
mean cosine of term $k$ with the net update; small or negative values mark a
large-magnitude term that mostly cancels against others.}
\label{tab:menu_usage_psi_ddot}
\begin{ruledtabular}
\begin{tabular}{lcc}
Vector Menu & Contribution $c_k$ & Alignment $\bar{\alpha}_k$ \\
\hline
$\dot{\Psi}_{\rm res}$ & $0.29$ & $0.35$ \\
$\mathcal{V}_0$ & $0.27$ & $0.46$ \\
$\nabla^2\Psi_{\rm res}$ & $0.14$ & $0.13$ \\
$\mathcal{V}_1$ & $0.12$ & $0.45$ \\
$\dot{\Psi}^{(1)}$ & $0.066$ & $0.33$ \\
$\nabla\mathcal{S}_0$ & $0.04$ & $0.024$ \\
$\mathbf{S}_\Psi\cdot\dot{\Psi}_{\rm res}$ & $0.036$ & $0.1$ \\
$\boldsymbol{\omega}_1$ & $0.034$ & $0.058$ \\
$\nabla\mathcal{S}_1$ & $0.017$ & $0.09$ \\
$\mathbf{S}_\pi\cdot\nabla^2\Psi_{\rm res}$ & $0.0082$ & $0.042$ \\
$\boldsymbol{\omega}_\pi\times\nabla^2\Psi_{\rm res}$ & $0.007$ & $0.048$ \\
$\dot{\Psi}^{(2)}$ & $0.0056$ & $0.048$ \\
$\mathbf{H}[\mathcal{S}_0]\cdot\dot{\Psi}_{\rm res}$ & $0.0054$ & $0.069$ \\
$\mathbf{S}_\pi\cdot\mathcal{V}_0$ & $0.003$ & $0.026$ \\
$\boldsymbol{\omega}_\pi\times\dot{\Psi}_{\rm res}$ & $0.0029$ & $0.038$ \\
\end{tabular}
\end{ruledtabular}
\end{table}

\textcolor{white}{super secret hidden text}

\newpage

\subsection{Physical Constraints \\ and Interpretability}
\label{app:interp}
\noindent 
In the main text we argued that the principal advantage of the Lagrangian Neural Cellular Automaton (LNCA) over existing convolutional emulators lies not in its raw accuracy but in the physical structure of its forward map. Here we substantiate that claim by examining the two properties that most sharply distinguish our architecture from the dense U-Net baselines: the strictly bounded rate at which information propagates across the Lagrangian lattice, and the transparency of the model's internal latent dynamics. We regard these not as aesthetic virtues but as properties of direct consequence for the physical fidelity of the emulator and for its utility as a differentiable forward model.

\begin{figure}[p] % [p] forces it onto a dedicated page/column float area
    \centering
    \begin{minipage}[c][\textheight][c]{\columnwidth}
        \centering
        \includegraphics[width=\columnwidth, height=0.95\textheight, keepaspectratio]{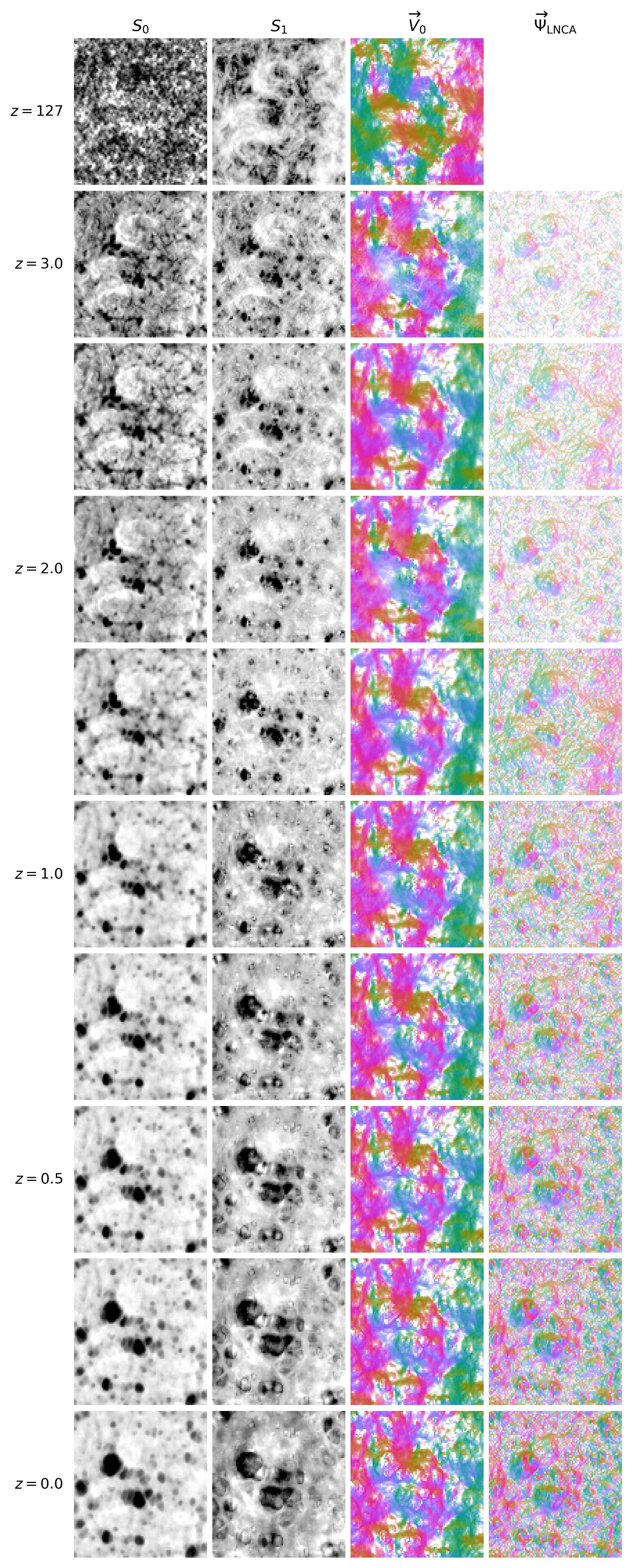}
        \caption{Time sequence of the latent scalar and vector fields, as well as the residual displacement field, over an inference rollout from $z=127 \rightarrow 0$. Each field is normalized independently to make feature more visible.}
        \label{fig:time_seq}
    \end{minipage}
\end{figure}

\subsubsection*{Transparency of the Latent Dynamics}
\noindent 
The first distinguishing property is the direct inspectability of the model's internal state dynamics. In a conventional convolutional emulator the intermediate activations are high-dimensional, frame-dependent feature maps with no fixed physical or geometric interpretation; they are difficult in general to visualize as anything like a field on the simulation volume, nor related to recognizable physical quantities. Our architecture is constructed so that every component of the node state retains a definite transformation character under $E(3)$ and a definite physical role throughout the rollout. The latent scalars $\mathcal{S}_i$ transform as invariants and accrue a memory of the local deformation history; the latent vectors $\mathcal{V}_i$ transform as genuine vectors and encode directional information such as the orientation of local infall and shear. Because each of these objects is a well-defined field on the Lagrangian lattice, the entire internal state sequence of the model may be rendered and examined.

Figure~\ref{fig:time_seq} presents such a rendering. We display central slices of the latent scalar and vector fields and the residual displacement field, at the initial conditions and at each of the target redshift snapshots $z \in \{127,\,3,\,2,\,1,\,0.5,\,0\}$. The evolution is highly structured and physically legible. At early times the latent fields are nearly featureless, reflecting the near-linearity of the displacement shortly after the initial conditions, where the Zeldovich approximation is itself an excellent description and the residual correction is correspondingly small. As the rollout proceeds toward $z=0$, coherent structure emerges in the latent channels along precisely the filaments, walls, and collapsing nodes of the cosmic web visible in the residual displacement field. The latent scalars develop pronounced amplitude in the regions of shell crossing, consistent with their interpretation as a learned record of where the Zeldovich description breaks down, while the latent vectors align with the local flow geometry and trace the coherent infall toward overdense regions.

\newpage

\newpage

\subsubsection*{Bounded Information Propagation}
\noindent 
The defining structural property of any cellular automaton is the locality of its update rule. In our construction, the state of a node $i$ at timestep $t+1$ depends only on the states of the $26$ Moore neighbors to which $i$ is permanently connected in the Lagrangian coordinate $q$. A single application of the update rule therefore transports information by exactly one cell, and over a rollout of $N_{\rm steps}$ substeps the causal influence of any node is confined to a Lagrangian ball of radius $N_{\rm steps}$ cells. This is the same finite propagation rate that motivates the boundary-cropping strategy described previously, where a margin of $N_{\rm steps}$ voxels is discarded to exclude nodes contaminated by inward-propagating stencil artifacts. 

%It stands in stark contrast to the U-Net architectures employed in previous field-level emulators \cite{Jamieson2025, He2019}, whose hierarchical downsampling and upsampling stages couple every point in the volume to every other in a single forward pass, imposing no bound on the rate of information propagation and admitting no notion of a causal cone.

%The physical relevance of this distinction is clear. Gravitational structure formation is itself a process of bounded information transport, in which the comoving distance over which a perturbation can causally influence its surroundings is set by the integrated particle displacement and remains modest (on the order a few $\mathrm{Mpc}/h$) even by $z=0$. An emulator whose receptive field is unbounded is free to construct spurious long-range correlations that have no counterpart in the underlying dynamics, and which can be suppressed only through careful regularization and large training volumes. Our architecture instead encodes locality directly into its computational graph, so that the emergent dynamical correlations are constrained, by construction, to respect a finite propagation speed on the lattice. Additionally, for later applications to field level inference, this property accelerates gradient backpropagation through the rollout, and the curvature of any likelihood surface defined through the emulator inherits the same locality.

What makes a bounded rule physically tenable is that the nonperturbative part of the map has finite support: the post-shell-crossing displacement of a Lagrangian patch is determined by the initial field within a radius comparable to the distance its material travels before collapse, while the perturbative, genuinely long-range component of the displacement is supplied analytically through the precomputed LPT templates described in \S\ref{sec:model_design} rather than being propagated by the automaton. The advantage of encoding locality in the computational graph is therefore not fidelity to a causal structure but the fact that gradient backpropagation through the rollout, and the curvature of any likelihood defined through it, inherit the same bounded support.

\begin{figure}[p]
  \centering
  \includegraphics[width=\linewidth]{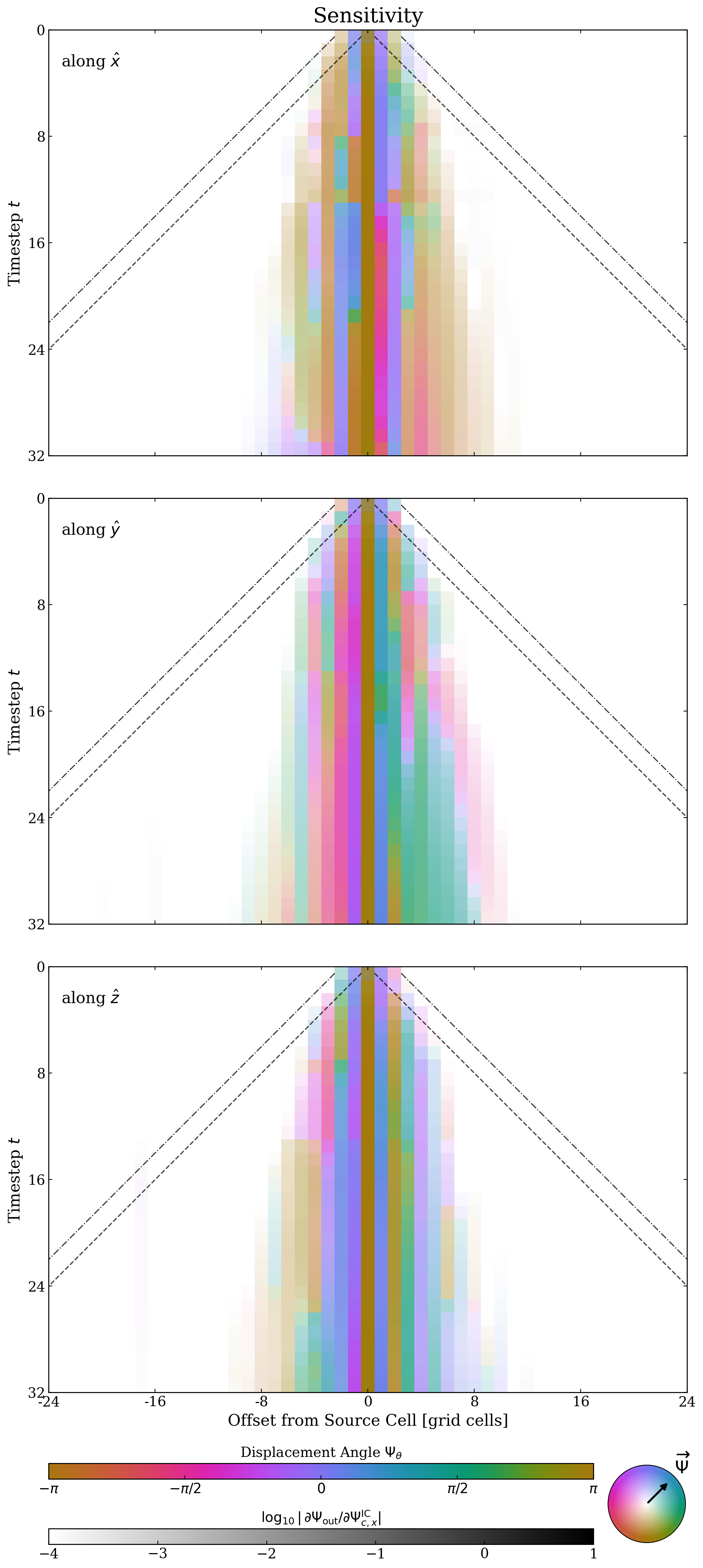}
  \caption{Sensitivity $\partial \Psi^t_x / \partial \Psi^{\rm IC}_0$ of the lightcone of cells downstream of a perturbation to a single cell in the $\hat{x}$ direction. Diagonal Lines represent the maximum rate of information transfer. Pixels have equal comoving spatial width and equal duration in $D(t)$.} 
  \label{fig:propagate}
\end{figure}

Figure~\ref{fig:propagate} demonstrates this property explicitly. We perturb the initial displacement of a single central cell by an infinitesimal amount along the $\hat{x}$ direction and, through forward-mode differentiation of the rollout, compute the resulting sensitivity $\partial \Psi^t_x / \partial \Psi^{\rm IC}_0$ of every downstream cell over all $32$ integration substeps. Stacking these responses against the integration time reveals the characteristic conical structure of the causal region: the perturbation influences only its immediate neighbors at early substeps, and the support of the response grows by at most one cell per step, tracing out a discrete light-cone on the Lagrangian lattice. The amplitude decays toward the surface of the cone and vanishes outside it. We emphasize that this bounded support follows from the localized receptive field rather than from any training penalty, and that it is a property of the learned operator alone, the LPT inputs on which it acts having been computed globally.

Taken together, these properties ensure that the LNCA does not merely produce an accurate final field but exposes a transparent account of \textit{how} that field was assembled, a window into the learned dynamics that is, to our knowledge, without precedent among field-level emulators of cosmic structure.

\newpage

\newpage

\bibliography{refs}

@misc{DESI2016,
  title={The {DESI} Experiment Part {I}: Science, Targeting, and Survey Design},
  author={{DESI Collaboration}},
  year={2016},
  eprint={1611.00036},
  archivePrefix={arXiv},
  primaryClass={astro-ph.IM},
  url={https://arxiv.org/abs/1611.00036}, 
}

@misc{Euclid2011,
  title={Euclid Definition Study Report},
  author={Laureijs, R and others},
  year={2011},
  eprint={1110.3193},
  archivePrefix={arXiv},
  primaryClass={astro-ph.CO},
  url={https://arxiv.org/abs/1110.3193},
}

@misc{LSST2009,
  title={LSST Science Book, Version 2.0},
  author={{LSST Science Collaboration}},
  year={2009},
  eprint={0912.0201},
  archivePrefix={arXiv},
  primaryClass={astro-ph.IM},
  url={https://arxiv.org/abs/0912.0201},
}

@article{Jasche2013,
  title={Bayesian physical reconstruction of initial conditions from large scale structure surveys},
  author={Jasche, Jens and Wandelt, Benjamin D},
  journal={Monthly Notices of the Royal Astronomical Society},
  volume={432},
  number={2},
  pages={894--913},
  year={2013},
  publisher={Oxford University Press}
}

@article{Wang2014,
  title={ELUCID-Exploring the Local Universe with the Reconstructed Initial Density Field. I. Hamiltonian Markov Chain Monte Carlo Method with Particle Mesh Dynamics},
  author={Wang, Huiyuan and Mo, HJ and Yang, Xiaohu and Jing, YP and Lin, WP},
  journal={The Astrophysical Journal},
  volume={794},
  number={1},
  pages={94},
  year={2014},
  publisher={IOP Publishing}
}

@article{Seljak2017,
  title={Generative model for cosmological large-scale structure},
  author={Seljak, Uro{\v{s}} and Aslanyan, Grigor and Feng, Yu and Modi, Chirag},
  journal={Journal of Cosmology and Astroparticle Physics},
  volume={2017},
  number={12},
  pages={009},
  year={2017},
  publisher={IOP Publishing}
}

@article{Modi2021,
  title={Cosmological inference with differentiable forward modeling},
  author={Modi, Chirag and Feng, Yu and Seljak, Uro{\v{s}}},
  journal={Journal of Cosmology and Astroparticle Physics},
  volume={2021},
  number={10},
  pages={035},
  year={2021},
  publisher={IOP Publishing}
}

@article{Buchert1992,
  title={Lagrangian theory of gravitational instability of Friedman-Lemaitre cosmologies and the'Zel'dovich approximation'},
  author={Buchert, Thomas},
  journal={Monthly Notices of the Royal Astronomical Society},
  volume={254},
  number={4},
  pages={729--737},
  year={1992},
  publisher={Oxford University Press}
}

@article{Bouchet1995,
  title={Perturbative Lagrangian approach to gravitational instability},
  author={Bouchet, F and Colombi, S and Hivon, E and Juszkiewicz, R},
  journal={Astronomy and Astrophysics},
  volume={296},
  pages={575},
  year={1995}
}

@article{Tassev2013,
  title={COLA: fast cosmological simulations using the Lagrangian perturbation theory with a particle-mesh algorithm},
  author={Tassev, Svetlin and Zaldarriaga, Matias and Eisenstein, Daniel},
  journal={Journal of Cosmology and Astroparticle Physics},
  volume={2013},
  number={06},
  pages={036},
  year={2013},
  publisher={IOP Publishing}
}

@article{He2019,
  title={Learning to predict the cosmological structure formation},
  author={He, Siyu and Li, Yin and Feng, Yu and Ho, Shirley and Ravanbakhsh, Siamak and Chen, Wei and P{\'o}czos, Barnab{\'a}s},
  journal={Proceedings of the National Academy of Sciences},
  volume={116},
  number={28},
  pages={13825--13832},
  year={2019},
  publisher={National Acad Sciences}
}

@article{Ramanah2019,
  title = {Painting halos from cosmic density fields of dark matter with physically motivated neural networks},
  author = {Kodi Ramanah, Doogesh and Charnock, Tom and Lavaux, Guilhem},
  journal = {Phys. Rev. D},
  volume = {100},
  issue = {4},
  pages = {043515},
  numpages = {18},
  year = {2019},
  month = {Aug},
  publisher = {American Physical Society},
  doi = {10.1103/PhysRevD.100.043515},
  url = {https://link.aps.org/doi/10.1103/PhysRevD.100.043515}
}

@article{Sanchez2020,
  title={Learning to simulate complex physics with graph networks},
  author={Sanchez-Gonzalez, Alvaro and Godwin, Jonathan and Pfaff, Tobias and Ying, Rex and Leskovec, Jure and Battaglia, Peter},
  journal={International conference on machine learning},
  pages={8459--8468},
  year={2020},
  organization={PMLR}
}

@article{Mordvintsev2020,
  author = {Mordvintsev, Alexander and Randazzo, Ettore and Niklasson, Eyvind and Levin, Michael},
  title = {Growing Neural Cellular Automata},
  journal = {Distill},
  year = {2020},
  note = {https://distill.pub/2020/growing-ca},
  doi = {10.23915/distill.00023}
}

@article{Grattarola2021,
  title={Learning graph cellular automata},
  author={Grattarola, Daniele and Livi, Lorenzo and Alippi, Cesare},
  journal={35th Conference on Neural Information Processing Systems},
  volume={34},
  pages={20983--20994},
  year={2021}
}

@article{Jamieson2025,
doi = {10.1088/1475-7516/2025/03/072},
url = {https://doi.org/10.1088/1475-7516/2025/03/072},
year = {2025},
month = {mar},
publisher = {IOP Publishing},
volume = {2025},
number = {03},
pages = {072},
author = {Jamieson, Drew and Li, Yin and Villaescusa-Navarro, Francisco and Ho, Shirley and Spergel, David N.},
title = {Field-level emulation of cosmic structure formation  with cosmology and redshift dependence},
journal = {Journal of Cosmology and Astroparticle Physics}
}

@article{Jamieson2022,
doi = {10.3847/1538-4357/acdb6c},
url = {https://doi.org/10.3847/1538-4357/acdb6c},
year = {2023},
month = {jul},
publisher = {The American Astronomical Society},
volume = {952},
number = {2},
pages = {145},
author = {Jamieson, Drew and Li, Yin and de Oliveira, Renan Alves and Villaescusa-Navarro, Francisco and Ho, Shirley and Spergel, David N.},
title = {Field-level Neural Network Emulator for Cosmological N-body Simulations},
journal = {The Astrophysical Journal}
}

@article{Villaescusa_Navarro_2020,
doi = {10.3847/1538-4365/ab9d82},
url = {https://doi.org/10.3847/1538-4365/ab9d82},
year = {2020},
month = {aug},
publisher = {The American Astronomical Society},
volume = {250},
number = {1},
pages = {2},
author = {Villaescusa-Navarro, Francisco and Hahn, ChangHoon and Massara, Elena and Banerjee, Arka and Delgado, Ana Maria and Ramanah, Doogesh Kodi and Charnock, Tom and Giusarma, Elena and Li, Yin and Allys, Erwan and Brochard, Antoine and Uhlemann, Cora and Chiang, Chi-Ting and He, Siyu and Pisani, Alice and Obuljen, Andrej and Feng, Yu and Castorina, Emanuele and Contardo, Gabriella and Kreisch, Christina D. and Nicola, Andrina and Alsing, Justin and Scoccimarro, Roman and Verde, Licia and Viel, Matteo and Ho, Shirley and Mallat, Stephane and Wandelt, Benjamin and Spergel, David N.},
title = {The Quijote Simulations},
journal = {The Astrophysical Journal Supplement Series}
}

@misc{Liu2025,
      title={KAN: Kolmogorov-Arnold Networks}, 
      author={Ziming Liu and Yixuan Wang and Sachin Vaidya and Fabian Ruehle and James Halverson and Marin Soljačić and Thomas Y. Hou and Max Tegmark},
      year={2025},
      eprint={2404.19756},
      archivePrefix={arXiv},
      primaryClass={cs.LG},
      url={https://arxiv.org/abs/2404.19756}, 
}

@article{Liu2025b,
  title = {Kolmogorov-Arnold Networks Meet Science},
  author = {Liu, Ziming and Tegmark, Max and Ma, Pingchuan and Matusik, Wojciech and Wang, Yixuan},
  journal = {Phys. Rev. X},
  volume = {15},
  issue = {4},
  pages = {041051},
  numpages = {25},
  year = {2025},
  month = {Dec},
  publisher = {American Physical Society},
  doi = {10.1103/4t7t-v19l},
  url = {https://link.aps.org/doi/10.1103/4t7t-v19l}
}

@misc{Ntampaka2021,
      title={The Role of Machine Learning in the Next Decade of Cosmology}, 
      author={Michelle Ntampaka and Camille Avestruz and Steven Boada and Joao Caldeira and Jessi Cisewski-Kehe and Rosanne Di Stefano and Cora Dvorkin and August E. Evrard and Arya Farahi and Doug Finkbeiner and Shy Genel and Alyssa Goodman and Andy Goulding and Shirley Ho and Arthur Kosowsky and Paul La Plante and Francois Lanusse and Michelle Lochner and Rachel Mandelbaum and Daisuke Nagai and Jeffrey A. Newman and Brian Nord and J. E. G. Peek and Austin Peel and Barnabas Poczos and Markus Michael Rau and Aneta Siemiginowska and Danica J. Sutherland and Hy Trac and Benjamin Wandelt},
      year={2021},
      eprint={1902.10159},
      archivePrefix={arXiv},
      primaryClass={astro-ph.IM},
      url={https://arxiv.org/abs/1902.10159}, 
}

@article{Ramanah2020,
    author = {Ramanah, Doogesh and Charnock, Tom and Villaescusa-Navarro, Francisco and Wandelt, Benjamin D},
    title = {Super-resolution emulator of cosmological simulations using deep physical models},
    journal = {Monthly Notices of the Royal Astronomical Society},
    volume = {495},
    number = {4},
    pages = {4227-4236},
    year = {2020},
    month = {05},
    issn = {0035-8711},
    doi = {10.1093/mnras/staa1428},
    url = {https://doi.org/10.1093/mnras/staa1428},
    eprint = {https://academic.oup.com/mnras/article-pdf/495/4/4227/33372045/staa1428.pdf},
}

@article{Tosone2021,
    author = {Tosone, Federico and Neyrinck, Mark C and Granett, Benjamin R and Guzzo, Luigi and Vittorio, Nicola},
    title = {muscle-ups: improved approximations of the matter field with the extended Press–Schechter formalism and Lagrangian perturbation theory},
    journal = {Monthly Notices of the Royal Astronomical Society},
    volume = {505},
    number = {2},
    pages = {2999-3015},
    year = {2021},
    month = {05},
    issn = {0035-8711},
    doi = {10.1093/mnras/stab1517},
    url = {https://doi.org/10.1093/mnras/stab1517},
    eprint = {https://academic.oup.com/mnras/article-pdf/505/2/2999/38626803/stab1517.pdf},
}

@article{Chartier2021,
    author = {Chartier, Nicolas and Wandelt, Benjamin and Akrami, Yashar and Villaescusa-Navarro, Francisco},
    title = {CARPool: fast, accurate computation of large-scale structure statistics by pairing costly and cheap cosmological simulations},
    journal = {Monthly Notices of the Royal Astronomical Society},
    volume = {503},
    number = {2},
    pages = {1897-1914},
    year = {2021},
    month = {02},
    issn = {0035-8711},
    doi = {10.1093/mnras/stab430},
    url = {https://doi.org/10.1093/mnras/stab430},
    eprint = {https://academic.oup.com/mnras/article-pdf/503/2/1897/36678136/stab430.pdf},
}

@article{Kaushal2022,
doi = {10.3847/1538-4357/ac5c4a},
url = {https://doi.org/10.3847/1538-4357/ac5c4a},
year = {2022},
month = {may},
publisher = {The American Astronomical Society},
volume = {930},
number = {2},
pages = {115},
author = {Kaushal, Neerav and Villaescusa-Navarro, Francisco and Giusarma, Elena and Li, Yin and Hawry, Conner and Reyes, Mauricio},
title = {NECOLA: Toward a Universal Field-level Cosmological Emulator},
journal = {The Astrophysical Journal}
}

@article{Doeser2024,
    author = {Doeser, Ludvig and Jamieson, Drew and Stopyra, Stephen and Lavaux, Guilhem and Leclercq, Florent and Jasche, Jens},
    title = {Bayesian inference of initial conditions from non-linear cosmic structures using field-level emulators},
    journal = {Monthly Notices of the Royal Astronomical Society},
    volume = {535},
    number = {2},
    pages = {1258-1277},
    year = {2024},
    month = {10},
    issn = {0035-8711},
    doi = {10.1093/mnras/stae2429},
    url = {https://doi.org/10.1093/mnras/stae2429},
    eprint = {https://academic.oup.com/mnras/article-pdf/535/2/1258/60481350/stae2429.pdf},
}

@article{Villaescusa-Navarro2021,
doi = {10.3847/1538-4357/abf7ba},
url = {https://doi.org/10.3847/1538-4357/abf7ba},
year = {2021},
month = {jul},
publisher = {The American Astronomical Society},
volume = {915},
number = {1},
pages = {71},
author = {Villaescusa-Navarro, Francisco and Anglés-Alcázar, Daniel and Genel, Shy and Spergel, David N. and S. Somerville, Rachel and Dave, Romeel and Pillepich, Annalisa and Hernquist, Lars and Nelson, Dylan and Torrey, Paul and Narayanan, Desika and Li, Yin and Philcox, Oliver and La Torre, Valentina and Maria Delgado, Ana and Ho, Shirley and Hassan, Sultan and Burkhart, Blakesley and Wadekar, Digvijay and Battaglia, Nicholas and Contardo, Gabriella and Bryan, Greg L.},
title = {The CAMELS Project: Cosmology and Astrophysics with Machine-learning Simulations},
journal = {The Astrophysical Journal}
}

@article{Gilpin2019,
  title = {Cellular automata as convolutional neural networks},
  author = {Gilpin, William},
  journal = {Phys. Rev. E},
  volume = {100},
  issue = {3},
  pages = {032402},
  numpages = {11},
  year = {2019},
  month = {Sep},
  publisher = {American Physical Society},
  doi = {10.1103/PhysRevE.100.032402},
  url = {https://link.aps.org/doi/10.1103/PhysRevE.100.032402}
}

@article{Wolfram1983,
  title = {Statistical mechanics of cellular automata},
  author = {Wolfram, Stephen},
  journal = {Rev. Mod. Phys.},
  volume = {55},
  issue = {3},
  pages = {601--644},
  numpages = {0},
  year = {1983},
  month = {Jul},
  publisher = {American Physical Society},
  doi = {10.1103/RevModPhys.55.601},
  url = {https://link.aps.org/doi/10.1103/RevModPhys.55.601}
}

@article{McAlpine2025,
  author  = {McAlpine, Simon and Jasche, Jens and Ata, Metin and others},
  title   = {The Manticore Project I: a digital twin of our cosmic neighbourhood from Bayesian field-level analysis},
  journal = {Mon. Not. Roy. Astron. Soc.},
  volume  = {540},
  pages   = {716},
  year    = {2025},
  eprint  = {2505.10682},
  archivePrefix = {arXiv},
  primaryClass  = {astro-ph.CO}
}

@article{Hahn2023,
doi = {10.1088/1475-7516/2023/04/010},
url = {https://doi.org/10.1088/1475-7516/2023/04/010},
year = {2023},
month = {apr},
publisher = {IOP Publishing},
volume = {2023},
number = {04},
pages = {010},
author = {Hahn, ChangHoon and Eickenberg, Michael and Ho, Shirley and Hou, Jiamin and Lemos, Pablo and Massara, Elena and Modi, Chirag and Moradinezhad Dizgah, Azadeh and Régaldo-Saint Blancard, Bruno and Abidi, Muntazir M.},
title = {SimBIG: mock challenge for a forward modeling approach to galaxy clustering},
journal = {Journal of Cosmology and Astroparticle Physics},
}

@misc{Nusser2026,
  author  = {Nusser, Adi},
  title   = {Bayesian Reconstruction of the Local Universe from 2MRS: Testing the Gravitational Flow with Cosmicflows-4},
  year    = {2026},
  eprint  = {2606.08593},
  archivePrefix = {arXiv},
  primaryClass  = {astro-ph.CO}
}

@article{Howlett2015,
  author  = {Howlett, Cullan and Manera, Marc and Percival, Will J.},
  title   = {L-PICOLA: A parallel code for fast dark matter simulation},
  journal = {Astron. Comput.},
  volume  = {12},
  pages   = {109--126},
  year    = {2015},
  eprint  = {1506.03737},
  archivePrefix = {arXiv},
  primaryClass  = {astro-ph.CO}
}

@article{BergerStein2019,
  author  = {Berger, Philippe and Stein, George},
  title   = {A volumetric deep Convolutional Neural Network for simulation of mock dark matter halo catalogues},
  journal = {Mon. Not. Roy. Astron. Soc.},
  volume  = {482},
  number  = {3},
  pages   = {2861--2871},
  year    = {2019},
  eprint  = {1805.04537},
  archivePrefix = {arXiv},
  primaryClass  = {astro-ph.CO}
}

@misc{spherex,
  author  = {Doré, Olivier and Bock, James and Ashby, Matthew and Capak, Peter and Cheng, Yan-Ting and de Putter, Roland and Eifler, Tim and others},
  title   = {Cosmology with the SPHEREX All-Sky Spectral Survey},
  year    = {2014},
  eprint  = {1412.4872},
  archivePrefix = {arXiv},
  primaryClass  = {astro-ph.CO}
}

@misc{roman,
  author  = {Spergel, David and Gehrels, Neil and Baltay, Charles and Bennett, David and Breckinridge, James and Donahue, Megan and Dressler, Alan and others},
  title   = {Wide-Field InfraRed Survey Telescope-Astrophysics Focused Telescope Assets WFIRST-AFTA 2015 Report},
  year    = {2015},
  eprint  = {1503.03757},
  archivePrefix = {arXiv},
  primaryClass  = {astro-ph.IM}
}

@misc{akeson2019,
  author  = {Akeson, Rachel and Armus, Lee and Bachelet, Etienne and Baker, Vanessa and Bartlett, Emma and Bean, Rachel and others},
  title   = {The Wide Field Infrared Survey Telescope: 100 Hubbles for the 2020s},
  year    = {2019},
  eprint  = {1902.05569},
  archivePrefix = {arXiv},
  primaryClass  = {astro-ph.IM}
}

@article{Hartl2026,
  author  = {Hartl, Benedikt and Levin, Michael and Pio-Lopez, L{\'e}o},
  title   = {Neural cellular automata: applications to biology and beyond classical AI},
  journal = {Phys. Life Rev.},
  volume  = {56},
  pages   = {94--108},
  year    = {2026},
  eprint  = {2509.11131},
  archivePrefix = {arXiv},
  doi     = {10.1016/j.plrev.2025.11.010}
}

@ARTICLE{Feng2016,
  author  = {{Feng}, Yu and {Chu}, Man-Yat and {Seljak}, Uro{\v{s}} and {McDonald}, Patrick},
  title   = {{FastPM}: a new scheme for fast simulations of dark matter and haloes},
  journal = {MNRAS},
  year    = {2016},
  volume  = {463},
  number  = {3},
  pages   = {2273},
  doi     = {10.1093/mnras/stw2123},
  eprint  = {1603.00476},
  archivePrefix = {arXiv},
  primaryClass  = {astro-ph.CO}
}

@ARTICLE{Dai2022,
  author  = {{Dai}, Biwei and {Seljak}, Uro{\v{s}}},
  title   = {Translation and rotation equivariant normalizing flow ({TRENF}) for optimal cosmological analysis},
  journal = {MNRAS},
  year    = {2022},
  volume  = {516},
  number  = {2},
  pages   = {2363},
  doi     = {10.1093/mnras/stac2010},
  eprint  = {2202.05282},
  archivePrefix = {arXiv},
  primaryClass  = {astro-ph.CO}
}

@ARTICLE{Dai2024,
  author  = {{Dai}, Biwei and {Seljak}, Uro{\v{s}}},
  title   = {Multiscale flow for robust and optimal cosmological analysis},
  journal = {Proc. Nat. Acad. Sci.},
  year    = {2024},
  volume  = {121},
  pages   = {e2309624121},
  doi     = {10.1073/pnas.2309624121},
  eprint  = {2306.04689},
  archivePrefix = {arXiv},
  primaryClass  = {astro-ph.CO}
}

@ARTICLE{Zhang2024,
  author  = {{Zhang}, Xinyi and {Lachance}, Philip and {Ni}, Yueying and {Li}, Yin and {Croft}, Rupert A.~C. and {Di Matteo}, Tiziana and {Bird}, Simeon and {Feng}, Yu},
  title   = {AI-assisted superresolution cosmological simulations III: time evolution},
  journal = {MNRAS},
  year    = {2024},
  volume  = {528},
  number  = {1},
  pages   = {281},
  doi     = {10.1093/mnras/stad3940},
  eprint  = {2305.12222},
  archivePrefix = {arXiv},
  primaryClass  = {astro-ph.CO}
}

@article{Li2022,
  author  = {Li, Yin and Lu, Libin and Modi, Chirag and Jamieson, Drew and Zhang, Yucheng and Feng, Yu and Zhou, Wenda and Kwan, Ngai Pok and Lanusse, Fran{\c c}ois and Greengard, Leslie},
  title   = {{pmwd}: A Differentiable Cosmological Particle-Mesh {$N$}-body Library},
  journal = {arXiv e-prints},
  year    = {2022},
  eprint  = {2211.09958},
  archivePrefix = {arXiv},
  primaryClass  = {astro-ph.CO}
}

@article{Nguyen2024,
  author  = {Nguyen, Nhat-Minh and Schmidt, Fabian and Tucci, Beatriz and Reinecke, Martin and Kosti{\'c}, Andrija},
  title   = {How Much Information Can Be Extracted from Galaxy Clustering at the Field Level?},
  journal = {Phys. Rev. Lett.},
  volume  = {133},
  number  = {22},
  pages   = {221006},
  year    = {2024},
  doi     = {10.1103/PhysRevLett.133.221006},
  eprint  = {2403.03220},
  archivePrefix = {arXiv},
  primaryClass  = {astro-ph.CO}
}

@article{Babic2025,
    doi = {10.1088/1475-7516/2025/11/066},
    url = {https://doi.org/10.1088/1475-7516/2025/11/066},
    year = {2025},
    month = {nov},
    publisher = {IOP Publishing},
    volume = {2025},
    number = {11},
    pages = {066},
    author = {Babić, Ivana and Schmidt, Fabian and Tucci, Beatriz},
    title = {Straightening the ruler: field-level inference of the BAO scale with LEFTfield},
    journal = {Journal of Cosmology and Astroparticle Physics},
}

@article{Porqueres2021,
  author  = {Porqueres, Natalia and Heavens, Alan and Mortlock, Daniel and Lavaux, Guilhem},
  title   = {Bayesian forward modelling of cosmic shear data},
  journal = {MNRAS},
  volume  = {502},
  number  = {2},
  pages   = {3035--3044},
  year    = {2021},
  doi     = {10.1093/mnras/stab204},
  eprint  = {2011.07722},
  archivePrefix = {arXiv},
  primaryClass  = {astro-ph.CO}
}

@article{Nguyen2020,
  author  = {Nguyen, Nhat-Minh and Jasche, Jens and Lavaux, Guilhem and Schmidt, Fabian},
  title   = {Taking measurements of the kinematic {Sunyaev-Zel'dovich} effect forward: including uncertainties from velocity reconstruction with forward modeling},
  journal = {JCAP},
  volume  = {2020},
  number  = {12},
  pages   = {011},
  year    = {2020},
  doi     = {10.1088/1475-7516/2020/12/011},
  eprint  = {2007.13721},
  archivePrefix = {arXiv},
  primaryClass  = {astro-ph.CO}
}

@article{Kwan2015,
doi = {10.1088/0004-637X/810/1/35},
url = {https://doi.org/10.1088/0004-637X/810/1/35},
year = {2015},
month = {aug},
publisher = {The American Astronomical Society},
volume = {810},
number = {1},
pages = {35},
author = {Kwan, Juliana and Heitmann, Katrin and Habib, Salman and Padmanabhan, Nikhil and Lawrence, Earl and Finkel, Hal and Frontiere, Nicholas and Pope, Adrian},
title = {COSMIC EMULATION: FAST PREDICTIONS FOR THE GALAXY POWER SPECTRUM},
journal = {The Astrophysical Journal}
}

@misc{thiele2022,
      title={Predicting the Thermal Sunyaev-Zel'dovich Field using Modular and Equivariant Set-Based Neural Networks}, 
      author={Leander Thiele and Miles Cranmer and William Coulton and Shirley Ho and David N. Spergel},
      year={2022},
      eprint={2203.00026},
      archivePrefix={arXiv},
      primaryClass={astro-ph.CO},
      url={https://arxiv.org/abs/2203.00026}, 
}

@ARTICLE{Bartlett2025,
       author = {{Bartlett}, Deaglan J. and {Chiarenza}, Marco and {Doeser}, Ludvig and {Leclercq}, Florent},
        title = "{COmoving Computer Acceleration (COCA): N-body simulations in an emulated frame of reference}",
      journal = {Astron. Astrophys.},
         year = 2025,
        month = feb,
       volume = {694},
          eid = {A287},
        pages = {A287},
          doi = {10.1051/0004-6361/202452217},
archivePrefix = {arXiv},
       eprint = {2409.02154},
 primaryClass = {astro-ph.IM},
       adsurl = {https://ui.adsabs.harvard.edu/abs/2025A&A...694A.287B}
}

\end{document}